\documentclass{aa} 
\usepackage{graphicx}
\begin{document}

\title{The 2-8 keV cosmic X-ray background spectrum as observed
with XMM-Newton\thanks{Based on observations with XMM-Newton, 
 an ESA science mission with instruments and contributions directly  funded by 
ESA member states and the USA (NASA)}}

\author{Andrea De Luca\inst{1,2} \and Silvano Molendi\inst{1}}

\institute{Istituto di Astrofisica Spaziale e Fisica Cosmica, 
Sezione di Milano  ``G.Occhialini'' - CNR 
v.Bassini 15, I-20133 Milano, Italy \and Universit\`a di Milano Bicocca, 
Dipartimento di Fisica, P.za 
della Scienza 3, 20126 Milano, Italy}
\offprints{A. De Luca,
\email{deluca@mi.iasf.cnr.it}}
   \date{Received ...; accepted ...}

\abstract{

We have measured the spectrum of the Cosmic X-ray Background (CXB) 
in the 2-8 keV range with the high throughput EPIC/MOS instrument onboard 
XMM-Newton. A large sample of high galactic latitude observations was used, 
covering a total solid angle of 5.5 square degrees. Our study is based on a 
very 
careful characterization and subtraction of the instrumental background, which 
is crucial for a robust measurement of the faintest diffuse source of the 
X-ray 
sky. The CXB spectrum is consistent with a power law having a photon index 
$\Gamma=1.41\pm0.06$ and a normalization of 2.46$\pm$0.09 photons cm$^{-2}$ 
s$^{-1}$ sr$^{-1}$ keV$^{-1}$ at 3 keV ($\sim$11.6 photons cm$^{-2}$ s$^{-1}$ 
sr$^{-1}$ keV$^{-1}$ at 1 keV), corresponding to a 2-10 keV flux of 
(2.24$\pm$0.16)$\times10^{-11}$ erg cm$^{-2}$ s$^{-1}$ deg$^{-1}$
(90\% confidence level, including the absolute flux calibration uncertainty). 
Our results 
are in excellent agreement with two of the most recent CXB measurements, 
performed with BeppoSAX  LECS/MECS data (Vecchi et al. 1999) and with an 
independent analysis of XMM-Newton EPIC/MOS data (Lumb et al. 2002), 
providing a very strong constrain to the absolute sky surface brightness in 
this 
energy range, so far affected by a $\sim$40\% uncertainty. Our measurement 
immediately implies that the fraction of CXB resolved by the recent deep X-ray 
observations in the 2-10 keV band is of $80\pm7$\% (1$\sigma$), suggesting the existence 
of 
a new population of faint sources, largely undetected within the current 
sensitivity limits of the deepest X-ray surveys.

}

\titlerunning{XMM-Newton Cosmic X-ray Background spectrum}
\authorrunning{De Luca \& Molendi}

\maketitle

\section{Introduction}
\label{intro}
The discovery of a diffuse Cosmic X-ray Background (CXB) radiation dates back 
to 
the birth of X-ray astronomy: it was a serendipitous result of the same rocket 
experiment which detected the first extra-solar X-ray source, Scorpius X-1 
(Giacconi et al. 1962). 
The problem of the nature of the CXB immediately became one of the most 
debated 
topics in astrophysics.

In the late 70's the first broad-band measurement of the CXB spectrum was 
obtained with the HEAO-1 satellite. In the 3-50 keV range the data were found 
to 
follow a thermal bremsstrahlung distribution (kT$\sim$40 keV), well 
approximated 
 below 15 keV by a simple power law with photon index $\Gamma\sim1.4$ 
(Marshall 
et al. 1980). 
In the same years several pieces of evidence (see e.g. the review by 
Fabian \& Barcons 1992) led to the understanding that the bulk of the CXB above 
the 
energy of $\sim$1 keV is extragalactic in origin.
COBE FIRAS observations (Mather et al. 1990) ruled out models based on diffuse 
emission from an hot intergalactic medium, strongly supporting the hypothesis 
that the CXB is made up from the integrated emission of faint discrete sources, 
with a dominant contribution from Active Galactic Nuclei (AGNs) (Setti \& 
Woltjer 1989).

This picture has been confirmed by the results from imaging X-ray 
observatories. 
Starting from the early observations with the Einstein satellite (Giacconi et 
al. 1979), coming to the recent deep surveys with Chandra (Moretti et al. 2003 
and references therein) and XMM-Newton 
(Hasinger at al. 2001), a higher and higher fraction of the CXB, up to 80-90\% 
in the overall 0.5-10 keV range, has been resolved into discrete sources, 
mainly 
obscured and 
unobscured AGNs. Indeed, the final solution of the origin of the CXB seems 
today 
to be quite close. The wealth of information coming from the deep pencil-beam 
X-ray surveys (Chandra Deep Field North, Brandt et al. 2001; Chandra Deep Field 
South, Giacconi et al. 2002), medium-deep wide angle X-ray surveys (HELLAS2XMM, 
Baldi et al. 2002) and their multiwavelength follow-up
campaigns, combined with synthesis models (e.g. Gilli et al. 2001), 
are defining the big picture, explaining which are the sources of the CXB and 
constraining their cosmological properties.

One of the main uncertainties involved in the problem is the CXB intensity 
itself. Since the HEAO-1 experiment, several measurements of the CXB spectrum 
have been obtained at energies below 10 keV. While the results on the spectral 
shape confirmed a power law with $\Gamma\sim1.4$, the normalization of the CXB 
remained highly uncertain as a consequence of large discrepancies (well beyond 
the statistical errors) among the different determinations. A difference as 
large as $\sim40$\% is found from the highest measured value (Vecchi et 
al. 1999 
using SAX data) to the lowest one (the original HEAO-1 experiment, Marshall et 
al. 1980). 

Barcons et al. (2000) pointed out that two different causes are required to 
explain the large scatter among the different determinations of the CXB 
intensity. First, cosmic variance: spatial variations of the CXB 
intensity are expected as a consequence of its discrete nature. This problem 
may be reduced by measurements covering large solid angles. Second, 
systematic errors, including cross-calibration differences, must have some 
role.
In any case, the measurements published after the analysis of Barcons et 
al. (2000), namely Lumb et al. (2002) with XMM-Newton EPIC and Kushino et 
al. (2002) with ASCA/GIS, while differing by $\sim$15\% only, either because of 
the small covered solid angle (Lumb et al. 2002), or because of large 
uncertainties in 
the stray light assessment (Kushino et al. 2002), do not allow to constrain the 
value 
of the CXB normalization to a much narrower range.

An uncertain value of the intensity represents a very severe limitation to any 
understanding of the ultimate nature of the CXB. Even a basic information such 
as the resolved fraction of the CXB cannot be firmly evaluated, leaving 
largely 
unsolved the problem of ``what is left'' beyond the detection limits of the 
deepest X-ray surveys: is there any fainter population still waiting to be 
resolved by even deeper observations? is there room for truly diffuse 
emission?

In this paper we present a new measurement of the CXB spectrum in the 2-8 keV 
range performed with the high throughput EPIC intrument onboard XMM-Newton. 
Our 
study is based on a large sample of high galactic latitude pointings for a 
total 
solid angle of $\sim$5.5 square degrees, reducing the effects of cosmic 
variance. Our analysis includes a very robust characterization 
of the instrumental background properties (including the issue of low-energy 
particle contamination - so far neglected in XMM-Newton data analysis), and 
particular care was devoted to the study of the possible sources of 
systematics.

The paper is organized as follows: in Sect.~\ref{nxbgeneral} we give an overview 
of the EPIC cameras instrumental background components and of the different 
approaches required for their correct subtraction. In Sect.~\ref{recipe} we 
describe in detail the algorithm which we used to extract the CXB spectrum 
starting from the raw EPIC data. In Sect.~\ref{results} we present our results; 
a 
detailed analysis of the possible sources of uncertainty is given in 
Sect.~\ref{syst}. In Sect.~\ref{discuss} our findings are discussed and compared 
to previous works. Our study of the EPIC instrumental background is quite 
technical and complex, but it may be useful for the study of extended sources 
such as  Clusters of Galaxies; for these reasons it is reported in detail in 
the Appendices~\ref{qnxb} and ~\ref{aqnxb}. In the figures throughout the paper 
error 
bars represent 1$\sigma$ uncertainties, unless otherwise specified.

\section{XMM/EPIC instrumental background issues}
\label{nxbgeneral}
The European Photon Imaging Camera (EPIC) instrument onboard XMM-Newton, 
consisting of two MOS CCD detectors (Turner et al. 2001) and a pn CCD 
camera (Str\"uder at al. 2001), has  appropriate characteristics  to 
study faint diffuse sources: it has an unprecedented collecting area ($\sim 
2500$ cm$^2$ @ 1 keV), a good spectral 
resolution ($\sim 6$\% @ 1 keV) and a large field of view (15 arcmin 
radius), over a rather broad energy range (0.1-12 keV). However 
the EPIC detectors were found to be affected on orbit by a 
rather high instrumental background noise (Non X-ray Background, NXB).
A correct characterization and subtraction of the NXB is the crucial step in 
order to get a robust measurement of the spectrum of the CXB, 
the faintest diffuse source in the X-ray sky.
In this work we will use data from the MOS cameras only. The pn 
detector, having different characteristics, will require a different 
approach.

\subsection{Overview of the different components of the MOS NXB}
\label{nxb}

The EPIC MOS internal background can be divided into two parts: a 
detector noise component and a particle-induced component.
The former is important only at low energies (below $\sim$0.4 keV) and 
will not be studied in this paper, since it is not a matter of concern 
for the measurement of the CXB in the 2-8 keV energy band; the latter 
dominates above 0.4 keV 
and therefore deserves a detailed characterization.

The signal generated by the interactions of particles with the 
detectors and with the surrounding structures has properties (temporal 
behaviour, spectral distribution, spatial distribution) largely 
depending on the energy of the impinging particles themselves. 

High energy particles (E$>$ a few MeV) generate a signal which is mostly 
discarded on-board on the basis of an upper energy thresholding and of 
a PATTERN analysis of the events (see e.g. Lumb et al. 2002). The unrejected 
part of this signal represent an important component of the MOS NXB. 
Its temporal behaviour is driven by the flux of energetic particles; 
its variability has therefore a time scale much larger than the length 
of a typical observation. We will refer to this NXB component as to 
the ``quiescent'' background.

Low energy particles (E$\sim$ a few tens of keV) accelerated in the 
Earth magnetosphere can also reach the detectors, scattering through 
the telescope mirrors. Their interactions with the CCDs generate 
events which are almost indistinguishable from valid X-ray photons and 
therefore cannot be rejected on-board. When a concentrated cloud of 
such particles is channeled by the telescope mirrors to the focal 
plane, a sudden increase of the quiescent count rate is observed. 
These episodes are known as ``soft proton flares'' since it is 
believed that the involved particles are protons of low energy (soft); the 
time scale is 
extremely variable, ranging from $\sim$100 s to several hours, while 
the peak count rate can be more than three orders of magnitude higher 
than the quiescent one. The extreme time variability is the 
fingerprint of this background component; it will be hereafter called 
the ``flaring'' NXB or ``Soft Proton'' (SP) NXB.

An additional component of background can be generated by a steady 
flux of low energy particles, reaching the detectors through the 
telescope optics at a uniform rate. No convincing evidence for the 
importance of such a NXB component have been so far presented (but see 
De Luca \& Molendi 2002 for a preliminary study) and its presence has 
been always neglected. In this work the existence of such quiescent low-energy 
particle background, as well as its impact on science, will be 
carefully studied.

\subsection{How to deal with the different MOS NXB components}
The standard result of an EPIC observation, after preliminary data 
processing, is an event list, basically containing the energy, the 
time of arrival and the position on the field of view of all of the 
collected photons. The list includes, besides good events generated by 
 photons from cosmic X-ray sources, spurious events due to the non X-ray
background. Noise events indeed represent the large majority in a 
typical blank sky pointing, when no bright sources are observed.

It is quite easy to identify the flaring background. Time 
variability is its signature, a light curve can immediately show the time 
intervals affected by an high background count rate. Such intervals 
are unusable for the analysis of faint diffuse sources and have to be 
rejected with the so-called Good Time Interval (GTI) filtering, which 
essentially consists in discarding all of the time interval having a count 
rate 
above a selected threshold.
The problem is particularly critical when the target of the observation is 
the CXB. A maximally efficient GTI filtering is required to study the 
faintest diffuse source of the sky: a good exposure time as high as 
possible is needed to maximize the statistics; conversely, even a 
low 
level of unrejected soft proton NXB could bias the measure of the CXB 
spectrum. The problem of GTI selection will be addressed in Sect.~\ref{gti}.

After the application of the GTI, a residual component of soft proton  
background may survive. This can result from  
different causes. For instance, low-amplitude flares, yielding little 
variations to the quiescent count rate, could be missed by the GTI 
threshold. Moreover, a slow time variability could hamper the 
identification of a ``flare'': in the most extreme case, a steady flow 
of particles impinging the detector during all of the duration of an 
observation would be almost impossible to identify by means of a time 
variability analysis. In such cases the unrejected NXB component could 
be revealed with a surface brightness analysis. As stated before, low 
energy particles are focused by the mirrors and therefore the spatial 
distribution of the induced NXB varies across the plane of the 
detectors.
We remember that a 
rather large portion of the MOS detectors is not exposed to the sky 
(hereafter called ``out Field Of View'', {\em out FOV}) and therefore 
neither cosmic X-ray photons nor low energy particle induced events are 
collected there\footnote{ Indeed, owing to the finite CCD transfer time,
a minor fraction of {\em in FOV} events is wrongly assigned to the {\em
out FOV} region as Out Of Time events. Such events correspond to $\sim$0.35\%
of the {\em out FOV} surface brightness and therefore they may be 
neglected to the aims of our work.}.
The study of the {\em out FOV} region allows to identify the 
observations affected by an anomalous low-energy particle NXB and to 
measure its level.
This issue will be discussed in Appendix~\ref{aqnxb}, where we also study 
the impact on science of such NXB component.

The final step required to remove the effects of NXB is to account for 
the quiescent component. Unfortunately, the quiescent NXB has no 
characteristic signatures and there are no recipes to separate good 
events from spurious events. A spectrum extracted from the event list 
obtained at this step (i.e. after GTI filtering and a check for the residual 
SP 
NXB) would be the superposition of the CXB and of the 
quiescent NXB. The only way to solve this problem is to get an independent 
measurement of the quiescent NXB spectrum. 
Its subtraction from the total (CXB$+$quiescent 
NXB) spectrum yields the pure 
CXB spectrum. The crucial problem is that the NXB spectrum, resulting 
from an independent measurement, must be  representative of the actual 
quiescent NXB which is present in a typical observation of the sky; 
otherwise, the determination of the CXB would be dramatically biased.
There are two ways to measure the quiescent NXB in the MOS. Firstly, 
through the 
analysis of the {\em out FOV} regions, where no X-ray photons or soft 
protons  
can reach the focal plane through reflections/scatterings by the telescope 
optics. Secondly, through the study of the observations with 
the filter wheel in {\em closed} position: in this configuration, an 
aluminium window prevents X-ray photons and low energy particles from 
reaching the detectors. The issue of a solid independent measure of the 
quiescent NXB is addressed in Appendix ~\ref{qnxb}, where we give the results 
of a long-term study of the quiescent background. 
Our analysis favours  the choice of 
the {\em closed} observations over the {\em out FOV}, as the former
provides an NXB spectrum better suited to study the CXB.

\section{The algorithm to measure the CXB spectrum}
\label{recipe}
This study is based on a rather large sample of MOS data including 
calibration, performance verification and granted time observations; 
public observations retrieved through the XMM-Newton Science Archive 
facility were also used. The initial dataset consist of (i) a compilation 
of (mostly) blank sky fields observations and (ii) a list of observations 
performed with the filter wheel in {\em closed} position.

We have developed an ad-hoc 
pipeline to perform the different steps of the analysis in an automated 
way. The implemented algorithm has the following main steps:

\begin{itemize}

\item For each individual observation (both sky fields and {\em closed}): 
preliminary data processing, event selection, GTI filtering; for the sky 
fields observations only: definition of the field of view (excision of 
the eventual bright target of the observation).
\item For sky fields observations: check for a residual soft proton 
contamination and definition  of the best dataset for the study of the 
CXB.

\item Stacking of the data (sky fields and {\em closed}), extraction of the 
source (sky) and background ({\em closed}) spectra corrected for vignetting.

\item Renomalization of the background spectrum, spectral analysis.

\end{itemize}

All of these points will be described in detail in the following sections. 
The automatic pipeline allowed to repeat the whole study varying the 
values for different parameters (from the flaring NXB threshold to the 
number of observations used) and to obtain a robust evaluation 
of the systematic errors involved, which will presented in  
Sect.~\ref{syst}. 
 
\subsection{Preliminary data selection and preparation}
\label{dataprepar}
We selected only high galactic latitude fields 
($|b|\ge28^{\circ}$). We avoided pointings towards the Magellanic Clouds, 
Cluster of Galaxies, as well as observations of very bright 
targets. The selected fields were observed between revolution number 57 
and revolution number 437. We retrieved the {\em closed} observations 
performed 
in the same time interval, between revolution 25 and 462.

The selected data were processed through the standard pipeline for 
event and energy reconstruction (emproc task in the XMM Science Analysis 
Software - SAS). In our long term project we used different 
releases of the SAS, from  v5.0 to v5.3.3. Different SAS versions 
are expected to give little differences to the aim of this work; we 
investigated this problem as a source of systematic errors in Sect.~\ref{sas}.

As stated in Sect.~\ref{nxb}, electronic noise is not a matter of 
concern in the study of the high energy (2-8 keV) CXB. However, hot  
pixels yielding a signal in the range of energy of interest have been 
occasionally observed. Bad pixels not uploaded for on-board rejection 
are automatically searched and discarded by the event reconstruction 
pipeline, but low-level flickering pixels may be missed. We developed 
an ad-hoc algorithm, based on the IRAF task cosmicrays, in order to 
identify and reject them.

Following standard prescriptions (see e.g. XMM-Newton Users' Handbook), we 
extracted good 
events with appropriate selections on the FLAG (the expression (FLAG \& 
0x766a0000)==0) allows to choose good events collected over the whole 
detector 
plane) and on the PATTERN (PATTERN$<$=12) parameters. 

We then applied
 a geometric mask  (a circle of 2.5 arcmin radius, including $>$98\% of
 the source counts)
 to exclude the eventual bright 
central target of the observation, to avoid biassing the determination of the 
CXB 
intensity, which is properly computed by integrating the contributions of {\em 
all} the resolved and unresolved {\em serendipitous} sources present in our 
sample of sky fields. 

For each observation, events from the {\em in FOV} and of the {\em out 
FOV} regions were separated and stored into two independent event lists. 
Events {\em in FOV} were selected with the expression  (FLAG \& 0x10000)==0, 
considering only a circle of 13.75 arcmin radius. Events from {\em out FOV} 
region were selected with  (FLAG \& 0x10000)!=0, with the  
constrain of an off-axis angle greater than 15 arcmin.
A further spatial mask was applied to the {\em out FOV} event list to 
discard 
a region in CCDs Number 2 and 7 
of both MOS1 and MOS2 cameras, 
where both X-ray photons and low energy 
particles can reach the detectors, 
possibly scattering through cuts in the camera 
body originally designed to accomodate a calibration source which was not 
installed.

\subsection{Flaring background screening}
\label{gti}
An efficient removal of the flaring background is a crucial step. A 
standard prescription (see e.g. Kirsch 2003) to identify high particle 
background time intervals is to study the light curve in the high 
energy (e.g. 10-12 keV) range, where the signal from cosmic sources is 
negligible.  However, it is 
observed that the flares generally turn on or off at different times 
at low and high energy (Lumb et al. 2002). 
Flares having a particularly soft spectrum
may be completely missed when the high energy range only is studied.

Since the study of the faint 
CXB requires a very efficient rejection of the flaring NXB, we decided 
to use the overall energy band 0.4-12 keV to search for the flares. We 
developed an algorithm to obtain an automated and homogeneous screening 
from the flaring NXB allowing for the identification of an ad-hoc count 
rate threshold for each observation. 

\begin{figure}
\centering

  \resizebox{\hsize}{!}{\includegraphics[angle=-90]{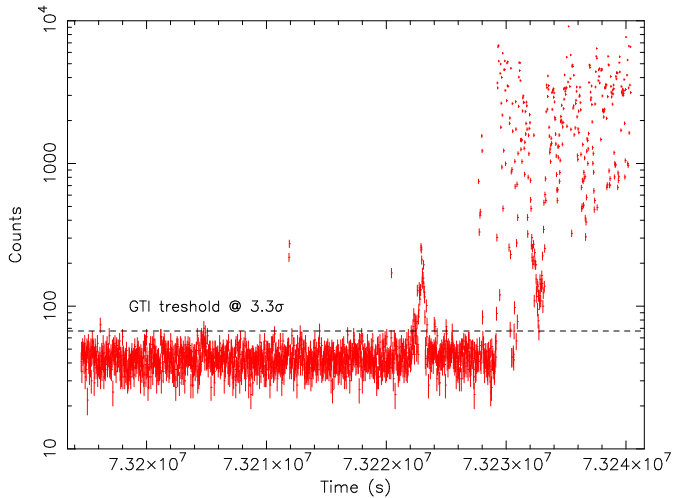}}
  \caption{ Light curve of a typical sky field observation. Events 
are extracted from the {\em in FOV} region in the energy range 0.5-12 keV; 
time bin is 30 s. The second part of the observation is affected by intense 
soft proton flares, the peak count rate is more than 200 times higher than 
the quiescent one. The selected threshold, in units of sigma of the 
quiescent count rate distribution (see text and Fig.~\ref{histo}), is marked 
with a dashed line.}
\label{lc}
\end{figure}

\begin{figure}
\centering

  \resizebox{\hsize}{!}{\includegraphics[angle=-90]{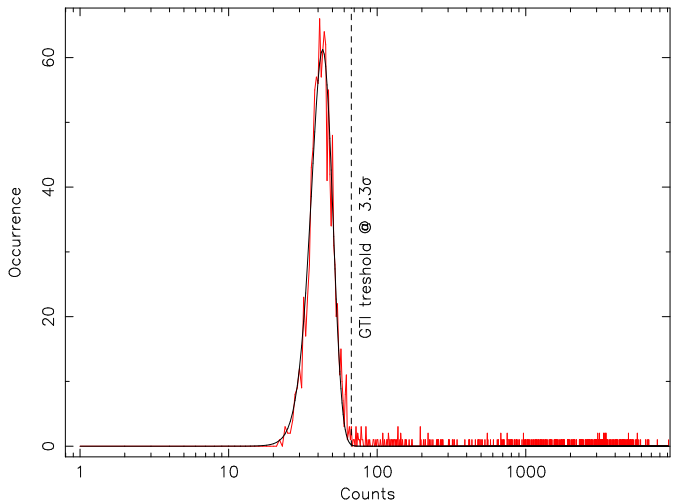}}
  \caption{  Histogram of the count rate distribution for a 
typical sky field observation. The corresponding light curve is shown in 
Fig.~\ref{lc}. The peak corresponds to the quiescent count rate Poisson 
distribution; points falling to the right correspond to the soft proton 
flares. The GTI threshold, in units of sigma of the quiescent count rate 
distribution, is marked with a dashed line. }
\label{histo}
\end{figure}

As a first step, a light curve in the 
0.4-12 keV   band is extracted with time bins of 30 s from the whole 
{\em in FOV} region. An histogram of the distribution of the counts is then 
built. The main peak in the histogram, corresponding to the poissonian 
distribution of the quiescent counts from the field, is identified and 
its position is computed by means of a simple fit with a Gaussian 
function (which represent an adequate approximation of the Poisson 
distribution for the observed mean number of counts per bin $\mu \sim 
40$). All of the time intervals which have a number of counts 
exceeding by more than 3.3$\sigma$ the mean are rejected. We discard also 
the time intervals corresponding to 0 counts. The chance occurrence 
probability 
of the ``0 counts event'' on 
the basis of the Poisson distribution is of order $e^{-40}$ or $\sim 
4\times10^{-18}$. The presence of a noticeable number of time bins 
with 0 counts is therefore to be ascribed to telemetry gaps or to 
other problems in the data flow; the corresponding time intervals have 
therefore to be rejected when computing the good exposure time.
The algorithm is applied independently on each of the datasets. 
The resulting good exposure time is computed summing the selected good 
time intervals, then applying the dead time correction.
In Fig.~\ref{lc} and ~\ref{histo} we show the typical light curve of a blank 
sky field (affected by intense SP flares in the last part) and the 
corresponding histogram of the counts/time bin. The adopted GTI threshold on 
the 
count rate is also shown in 
both cases.

The flaring background signal, as expected, is not seen either in the {\em 
out 
FOV} region in the sky fields observations, or in the {\em closed} 
observations. 
Nevertheless, the GTI identified for each sky field
observation was applied also to the corresponding event list for the 
{\em out FOV} region of the detectors, in order to allow for a coherent 
comparison of the datasets. For consistency, a GTI filtering using the same 
prescription (for both the extraction of the light curve and the selection 
of the threshold) was also performed on the {\em closed} observations, 
yielding 
in any case the rejection of a negligible part of the exposure time.

In Sect.~\ref{systgti} we will study how a different choice for the GTI 
threshold 
can affect the determination of the CXB spectral parameters.

\subsection{Check for residual soft proton background}
\label{check}
A component of soft proton background may survive after GTI screening:  
two peculiar phenomenologies of flares cannot be 
properly identified by our automated algorithm. This is the case of (i) 
extreme 
variability in the light curve, 
(ii) 
conversely, very slow time variability of the soft proton flux. In both 
cases, the peak in the 
histogram of the count rates (see previous section) has a 
significant contribution from soft proton NXB. This could introduce a bias 
in the measure of the faint CXB spectrum. 

We have developed a simple 
diagnostic to identify the observations which are contaminated by such a soft 
proton NXB 
component. The ratio $R$ between the surface brightness {\em in 
FOV} ($\Sigma_{IN}$) and the surface brightness {\em out FOV} ($\Sigma_{OUT}$) 
in the range 8-12 keV is used to identify an anomalous   
NXB affecting only the {\em in FOV} region (as expected in the case of a 
residual soft 
proton component, owing to the focusing of the low energy particles by the 
telescope optics). 
The details of our study of the residual soft proton NXB component, 
demonstrating its impact on the measure of the CXB spectrum, are presented 
in Appendix ~\ref{aqnxb}. 

The spectral shape of the contaminating NXB was 
found to vary in an unpredictable way from observation to observation.
To get a robust measure of the CXB, we  decided therefore to discard the most 
affected observations. 
Particular care was then devoted to the selection of the best observations, 
free from significant residual soft proton contamination, setting an 
appropriate 
threshold ($R_{max}=1.3$) on the value of the ratio of the surface 
brightnesses 
$\Sigma_{IN}/\Sigma_{OUT}$ (see Sect.~\ref{aqxbsyst} for a study of the 
possible involved systematics).
As a result, we rejected 9 observations out of 51 for the MOS1 and 6 out of 
49 for the MOS2, corresponding to the 12\% and 10\% of the total exposure 
time, respectively. 


\subsection{Stacking of the data}
\label{merging}
We merged all the selected observations using the SAS task evlistcomb (we are 
interested in the detector coordinates only) to obtain four event 
lists, corresponding to the {\em in FOV} and the {\em out FOV} data for both 
the 
sky fields and the {\em closed} observations. 

For each merged event list we built an appropriate exposure map, computing 
the total exposure time corresponding to each position in detector 
coordinates. The resulting exposure map for the sky fields event list {\em in 
FOV} is not uniform due to the rejection of the central bright sources 
which were the original targets of the individual observations and to the 
presence of a few observations in small window mode. 
Conversely, the exposure map for the {\em closed} observations is of course
 flat.

\subsection{Extraction of the spectra corrected for vignetting}
\label{vfcorr}
The decrease of the effective areas as a function of increasing off-axis 
angle causes a loss of flux which is known as vignetting. To correct for 
this 
effect, we developed an 
algorithm based on a photon weighting method, similar to the CORRECT 
algorithm implemented in the EXSAS software for ROSAT (Zimmermann et 
al. 1998). Each photon having an energy $E_j$ belonging to the I$^{th}$ 
spectral channel ($E(I)<E_j<E(I+1)$) falling at the  position 
($x_j$,$y_j$) in detector coordinates gives a contribution 
$$\frac{A_{eff}(0,0,E_j)}{A_{eff}(x_j,y_j,E_j)} \frac{1}{t(x_j,y_j)}$$ to 
the count rate in the spectral channel I; in the 
previous relation $A_{eff}(0,0,E_j)$ and $A_{eff}(x_j,y_j,E_j)$ are the 
effective areas at energy $E_j$ on-axis and at the position ($x_j,y_j$) 
respectively; the exposure time $t(x_j,y_j)$ is read from the exposure 
map (see previous section). In this way the extracted spectrum is 
automatically corrected for the vignetting. Spectra were accumulated for both 
PATTERN 0-12 events (the standard selection, see Kirsch 2003) and PATTERN 0 
only 
(single pixel events) for a consistency check. 

The background spectrum from the merged {\em closed} event file was 
obtained using the same method. The NXB is not vignetted; however, since 
the correction is applied to the total (CXB+NXB) spectrum from the merged 
sky event list, the same correction to the (pure NXB) spectrum, extracted 
from the same detector region, is required. This photon weighting method has 
the 
advantage 
of allowing for an exact correction of the vignetting effect, accounting 
for a non-uniform exposure map as well as for spatial variations in the 
NXB.

\subsection{Renormalization of the NXB spectrum} 
\label{renorm}
A renormalization is needed to account for the different intensities 
of the NXB in the sky fields and in the {\em closed} observations due to the 
non-simultaneity of the
measurement.
To compute the renormalization factor, taking advantage of the very high 
statistic, we used the energy range 10-11.2 keV where no fluorescence 
lines (possibly having different time behaviour wrt. the continuum, see 
Appendix ~\ref{qnxb}) are present and where the expected CXB signal is 
negligible, being of order 0.2\%. The ratio of the count rates in the 
total (CXB+NXB) and (pure NXB) spectra is computed and used as 
renormalization factor for the NXB spectrum. In Sect.~\ref{nxbreno} we will 
study the effects of the uncertainty in this operation.

\subsection{Spectral analysis}
\label{analysis}
The spectra from the merged sky fields were rebinned by a factor of 10 to 
obtain 
a good statistic in each energy channel after background subtraction. The 
adopted vignetting correction method requires the use of the on-axis 
redistribution 
matrices\footnote{ The ground calibration campaign demonstrated that the Quantum 
Efficiencies of the different CCDs are uniform above $\sim$0.3 keV (the 
differences were 
measured to be $<$2\% at 2 keV, 
Vercellone 2000). In our analysis we account for possible systematics affecting 
the 
absolute flux measurement as discussed in Sect.~\ref{absolute}.}
 and effective area files. For each camera, we used an exposure-weighted 
effective area file computed as a linear combination of the  MOS1/2 
effective areas 
provided by the XMM-Newton Science Operation Centre for the different optical 
filters, the weights being 
the fractions of the total exposure time corresponding to each filter. 

The 2-8 keV range was selected for the analysis. Lower energies were not used 
to 
avoid (i) contaminations by the soft galactic component of the CXB (emerging 
below $\sim$1 keV) and (ii) possible artefacts due to an imperfect subtraction 
of the bright Al-K and Si-K fluorescence lines (in the 1-2 keV range - see 
App. 
~\ref{qnxb}). Above 8 keV the collected CXB signal is marginal.

The spectral analysis was performed within XSPEC v11.0.
The spectral model was a simple absorbed power law. The interstellar 
absorption 
$N_H$ was fixed to the exposure-weighted average of the values of the 
selected fields. We added to the model multiple gaussian lines to account for 
possible 
differences in intensity of the brightest fluorescence lines (Cr, Mn, Fe in 
the 
5-7 keV spectral range) in the sky and in the {\em closed} observations. Their 
energies and FWHM  (which are constant across the detector plane, see 
Sect.~\ref{spatial})
were fixed to the values computed using the {\em out FOV} data (sky and 
{\em closed} 
observations yielded fully consistent results). 

To minimize the correlation between the CXB spectral parameters (photon index 
and intensity), the 
normalization of the power law was evaluated at the barycentre of the selected 
energy range (see Ulrich \& Molendi 1996), which in our case is found to lie 
at 
$\sim$3 keV.  MOS1 and 
MOS2 data were studied both separately and in a symultaneous fit.

\section{Results}
\label{results}
The final dataset includes 42 sky fields for the MOS1 camera and 43 for the 
MOS2. The total exposure time is of $\sim$1.15 Ms per camera. The solid angle 
covered by the data, summing the contribution of each observation (accounting 
for the differences in field of view due to the readout mode or to the 
excision 
of the central target) is of $\sim$5.5 square degrees (34 different pointing 
directions) per camera.
The {\em closed} data amount to $\sim$430 ksec per camera.

\begin{figure}
\centering

  \resizebox{\hsize}{!}{\includegraphics[angle=-90]{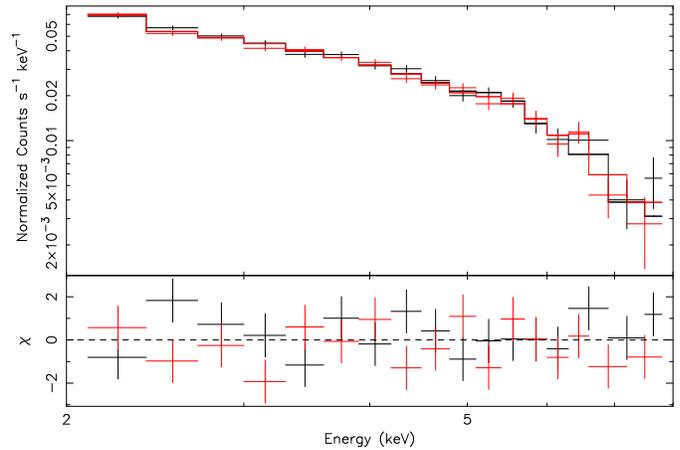}}
  \caption{  The cosmic X-ray background spectrum in the 2-8 keV 
range is displayed, folded with the instrumental response. MOS1 data are 
represented in red, MOS2 in black. The best fit model is overplotted. The 
lower 
panel shows the residuals in units of sigma.}
\label{bestfit}
\end{figure}

The spectrum of the cosmic X-ray background as seen by the MOS instruments is 
shown in Fig.~\ref{bestfit}. We note that the CXB signal is very low if 
compared 
to the quiescent NXB, accounting for only $\sim$20\% of the counts in the  
(vignetting-corrected) spectrum from the sky field merged dataset in the 2-8 
keV 
range.

The two cameras yield fully consistent results within the statistical 
uncertainties (see Table ~\ref{tabresults}). A symultaneous fit to the data 
($\chi^{2}_{\nu}=1.15$, 72 d.o.f.) yields a photon index of 1.41$\pm$0.04 and 
a 
normalization of $2.647\pm0.038$ photons cm$^{-2}$ s$^{-1}$ sr$^{-1}$ 
keV$^{-1}$ 
at 3 keV (to be corrected for the stray light, i.e. the contribution to the 
collected flux due to photons coming from out-of-field angles). The quoted 
uncertainties 
are the statistical errors at the 90\% confidence level for a single 
interesting 
parameter. 

\begin{table*}[h]
\begin{center}
\begin{tabular}{cccc} \hline \hline
Instrument & Photon Index &   Normalization & $\chi^{2}_{\nu}$/d.o.f.  \\ 
\hline
MOS1   & 1.43$\pm$0.07 & 2.62$\pm$0.05 & 1.16/35  \\ \hline
MOS2      & 1.38$\pm$0.07 & 2.68$\pm$0.05  &  1.17/35   \\  \hline
MOS1+MOS2      & 1.41$\pm$0.04 & 2.647$\pm$0.038 & 1.15/72 \\ \hline \hline
\end{tabular}
\end{center}
\caption{\label{tabresults} Results of the spectral analysis on the CXB for 
the 
MOS cameras. The normalization is expressed in photons cm$^{-2}$ s$^{-1}$ 
sr$^{-1}$ keV$^{-1}$ at 3 keV (not corrected for the stray light 
contribution). 
The quoted uncertainties are purely statistical errors at the 90\% confidence 
level for a single interesting parameter.} 

\end{table*}

A careful study of the possible sources of errors led us to compute the 
overall 
uncertainty (systematics included) to be of 4\% for the photon index and of 
3.5\% for the normalization. The details of the error analysis are presented 
in 
the next section. 

After correcting for the stray light (see Sect.~\ref{stray}), the MOS results 
on 
the 2-8 keV CXB spectrum are:

$$\Gamma\,=\,1.41\pm0.06$$
$$N\,=\,2.462\pm0.086$$ 

where the normalization $N$ is expressed in photons cm$^{-2}$ s$^{-1}$ sr$^{-1}$ 
keV$^{-1}$ at 3 keV.

The resulting flux in the 2-10 keV energy range is of 
(2.24$\pm$0.16)$\times10^{-11}$ erg   
cm$^{-2}$ s$^{-1}$ deg$^{-2}$. 
 The error (90\% confidence) includes also an extra 5\% uncertainty as an 
estimate
of the absolute flux calibration accuracy of the MOS cameras.
To ease a comparison with previous works, the 
corresponding normalization at 1 keV is of $\sim$11.6 photons cm$^{-2}$ 
s$^{-1}$ 
sr$^{-1}$ keV$^{-1}$.

\section{Analysis of errors}
\label{syst}
A very careful evaluation of the possible sources of error, including 
systematics, is of 
paramount importance in the measure of a faint source such as the CXB.
In the following sections the quoted errors on the spectral parameters are at 
90\% confidence level for a single interesting parameter.

\subsection{Effect of the use of different SAS versions}
\label{sas}
As stated in Sect.~\ref{dataprepar}, in this work different SAS releases 
(namely v5.0, v5.3.0, 5.3.3) were used to perform the preliminary 
processing of the data.  We processed $\sim$ 100 ksec per camera of {\em 
closed} 
observations with the SAS v5.3.3 to be compared with the corresponding 
datasets in our compilation processed through SAS v5.0. 
The analysis and the extraction of the spectra were performed as described 
in Sect.~\ref{recipe}. 
A linear fit to the ratio of the two spectra in the 
range 2-12 keV was found to be fully consistent with a constant equal to 1, 
the slope being statistically null.
Nevertheless, as a further step we 
used the computed (90\% confidence) limits on the slope to modify the final 
quiescent NXB spectrum (i.e. the one used in 
Sect.~\ref{results}); such ``distorted'' spectrum was used to study the 
CXB, running the second part of our pipeline. We obtained a variation of 
$\sim$2\% in the best fit photon index value, and a variation 
of $\sim$1.5\% in the normalization. Such values, comparable to the 
statistical uncertainties (see Sect.~\ref{results}), represent a very 
conservative estimate  (indeed, a worst case) 
of the possible systematics associated with the use 
of different SAS versions.

\subsection{Effect of different thresholds for soft proton screening}
\label{systgti}
We took advantage of our automated pipeline to study the effects of 
different prescriptions for the GTI filtering (see Sect.~\ref{gti}) 
directly on the best fit parameters of the CXB, simply running the whole 
pipeline several times, using different values for the GTI selection 
parameters. 
While different recipes to build the light curve are generally less 
efficient, we investigated the effects of a different choice for 
the count rate threshold. 
This is indeed a crucial problem. Lumb et al. (2002) reported a variation of 
the 
cosmic 
background spectral parameters as a results of different flare screenings, but 
they did not present any systematic study of this effect.

\begin{figure}
\centering

  \resizebox{\hsize}{!}{\includegraphics[angle=-90]{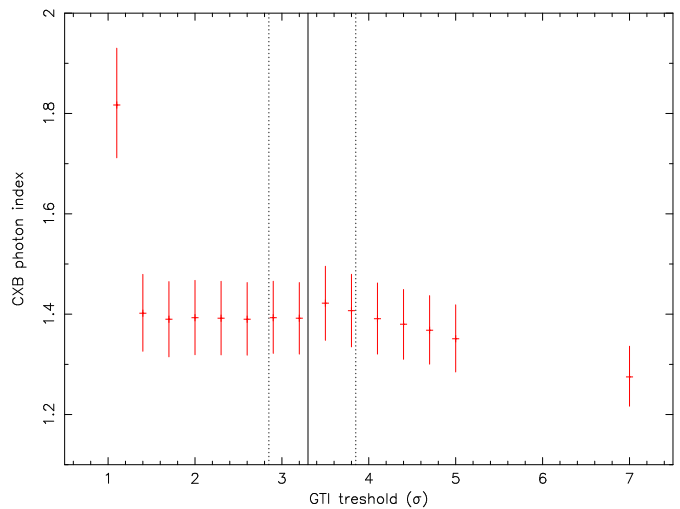}}
  \caption{ The best fit photon index of the CXB is 
plotted as a function of the selected GTI threshold (in units of sigma of the 
quiescent count rate distribution - see Sect.~\ref{gti}). Errors 
(at 90\% confidence level for a single interesting parameter) include the 
uncertainty on the renormalization of the quiescent NXB spectrum. The adopted 
3.3$\sigma$ threshold is marked by the vertical solid line. No significant 
variations in the CXB photon index are obtained as a result of different 
choices of the GTI threshold within 
the range $\sim2.9-3.8\sigma$, marked by the dotted vertical lines.}
\label{gtisystslope}
\end{figure}

\begin{figure}
\centering

  \resizebox{\hsize}{!}{\includegraphics[angle=-90]{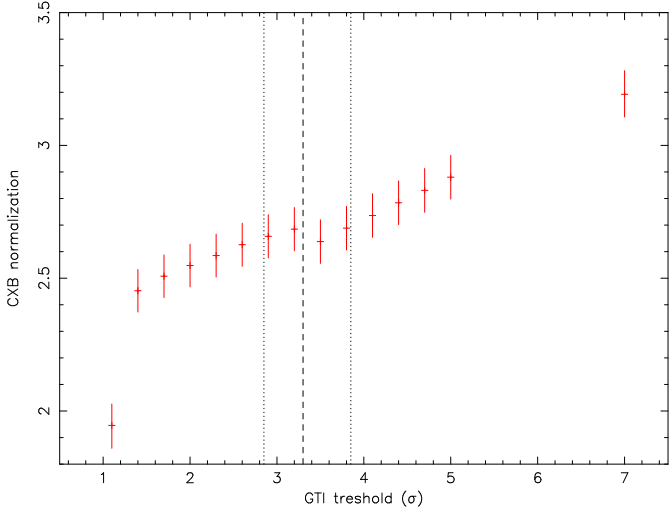}}
  \caption{ Same as Fig.~\ref{gtisystslope}, the best fit 
CXB normalization is shown. Units are photons cm$^{-2}$ s$^{-1}$ sr$^{-1}$ 
keV$^{-1}$ at 3 keV. The contributions from out-of-field scattered light have 
not been subtracted. Errors (at 90\% confidence level for a single interesting 
parameter) include the uncertainty on the renormalization 
of the quiescent NXB spectrum. The adopted GTI threshold is marked by the 
vertical dashed line. It is evident that no significant variations are 
obtained as a result of small changes of the selected GTI threshold (within 
the range $\sim2.9-3.8\sigma$, marked by the dotted vertical lines). See 
text for further details.}
\label{gtisystnorm}
\end{figure}

The results for the photon index and the 
normalization of the CXB as a function of the threshold (in units of sigma 
from the average count rate) are shown in Fig.~\ref{gtisystslope} and 
Fig.~\ref{gtisystnorm}. It is evident that the selected 3.3$\sigma$ level 
represent an optimal threshold.
Little changes in the adopted value (in both 
directions) have essentially no effects on the best fit CXB parameters. 
The choice of significantly lower values implies the rejection of positive 
random fluctuations from the good count rate, lowering the inferred 
normalization; conversely, the choice of significantly higher values 
implies the inclusion of a fraction of soft proton NXB which increase the 
derived CXB normalization. Within the ``good'' range for 
the GTI threshold identified in this study ($\sim2.9\sigma-3.8\sigma$), the  
variations of the photon index and normalization of the CXB spectrum are 
always smaller than the statistical uncertainties. We therefore consider any 
systematic error associated with the GTI thresholding to be negligible.

\subsection{Effect of a different threshold for the residual soft proton 
NXB}
\label{aqxbsyst}
\begin{figure}
\centering

  \resizebox{\hsize}{!}{\includegraphics[angle=-90]{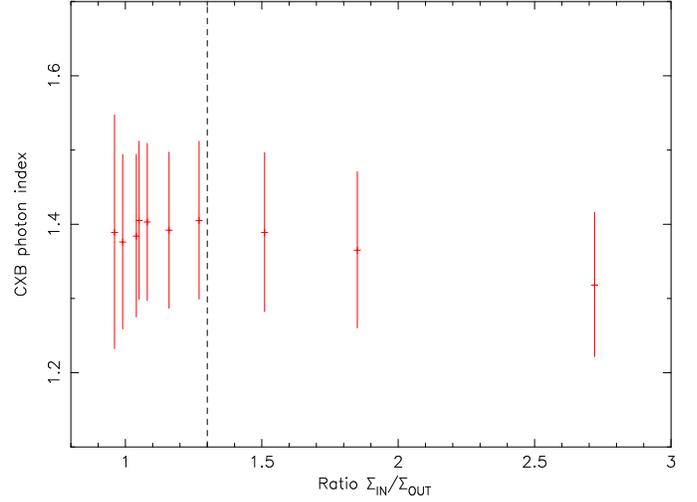}}
  \caption{ Best fit CXB photon index as a function of 
the maximum allowed ratio for the 8-12 keV surface brightness 
$\Sigma_{IN}/\Sigma_{OUT}$, a probe of the presence of residual soft proton 
NXB. The MOS1 case is shown (MOS2 is very similar). Errors (at 90\% confidence 
level 
for a single interesting parameter) include the 
uncertainty on the renormalization of the quiescent NXB spectrum. No 
significant 
changes 
are obtained. The case of the CXB normalization is markedly different (see 
Fig.~\ref{norm_vs_ratio}).}
\label{slope_vs_ratio}
\end{figure}

\begin{figure}
\centering

  \resizebox{\hsize}{!}{\includegraphics[angle=-90]{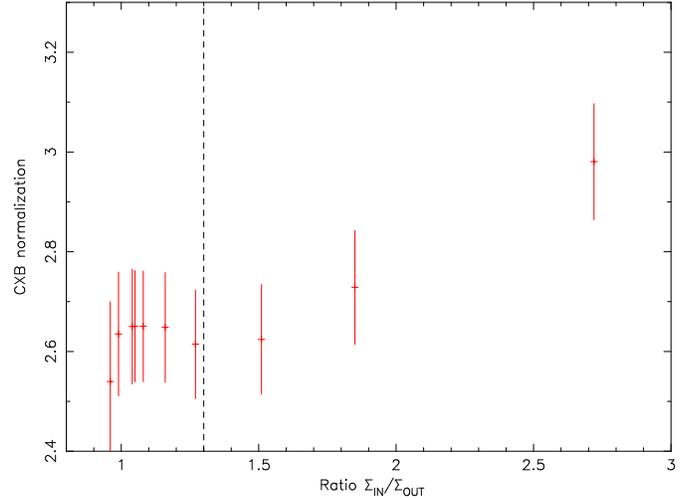}}
  \caption{ Same as Fig.~\ref{slope_vs_ratio}, the best 
fit CXB normalization (in units of photons cm$^{-2}$ s$^{-1}$ sr$^{-1}$ 
keV$^{-1}$ at 3 keV) is plotted here. The stray light correction have not been 
applied. Errors (at 90\% confidence level for a single interesting parameter) 
include 
the uncertainty on the renormalization of the 
quiescent NXB spectrum. When dataset having large values of 
the ratio $\Sigma_{IN}/\Sigma_{OUT}$ are included, the normalization is 
systematically higher owing to the presence of a significant residual SP NXB 
component (see text). The adopted maximum allowed ratio ($R_{MAX}=1.3$) is 
marked by the vertical line. No significant changes are obtained within 
small changes of $R_{MAX}$. See Appendix ~\ref{aqnxb} for further details.}
\label{norm_vs_ratio}
\end{figure}

As shown in Appendix~\ref{impact}, the residual soft proton background may 
introduce a strong bias in the measure of the CXB spectral parameters. 
It is therefore very important to check if our recipe to define a good, 
non-contaminated dataset on the basis of the  ratio of surface brightnesses 
$\Sigma_{IN}/\Sigma_{OUT}$ (see Sect.~\ref{check} and Appendix 
~\ref{aqnxb}) is a possible source of systematics.
We studied directly the variation of the CXB spectral parameters as a 
function of the threshold $R_{max}$ on the ratio $\Sigma_{IN}/\Sigma_{OUT}$.
For each selected value of $R_{max}$ (we explored the range $\sim$0.95-2.7), 
the 
datasets {\em below} 
the threshold were merged and the CXB spectrum extracted and analyzed 
running our automatic pipeline. The 
computed best fit CXB
 photon index and normalization are plotted as a function of 
the threshold $R_{max}$ in Fig.~\ref{slope_vs_ratio} and 
~\ref{norm_vs_ratio}, respectively. While the value  of the photon index does 
not change significantly as a function of $R_{max}$, in the case of the 
normalization a peculiar trend is seen and can be 
easily interpreted. When the threshold $R_{max}$ is high ($>$1.5), the 
contributions of the contaminated observations is important. Decreasing 
values of $R_{max}$ yield therefore a significant variation in the CXB 
normalization 
 as a result of the rejection of an increasing number of bad 
observations. The trend disappears and the normalization become 
stable when the contaminated observations have been discarded. Further 
decreases of $R_{max}$ lead to the exclusion of good observations with 
positive fluctuations of the ratio $\Sigma_{IN}/\Sigma_{OUT}$. The break 
occurs 
for $R_{max}$ $\sim$1.5. We assume conservatively $R_{max}$=1.3 to define 
our 
final dataset to study the CXB.
An inspection of Fig.~\ref{slope_vs_ratio} and ~\ref{norm_vs_ratio} 
shows that for slightly different choices of $R_{max}$ in the ``good'' range 
$\sim1.0\div1.4$ the variations of the best fit CXB slope and 
normalization are small with respect to the statistical uncertainties. 
We conclude therefore that our approach for identifying and rejecting 
contaminated observations is robust and free from systematics.

 We note in addition that a new break is observed in Fig.~\ref{norm_vs_ratio} 
for $R_{max} \sim$0.95.
This occurs when the threshold $R_{max}$ is below the average value of the 
ratio 
$\Sigma_{IN}/\Sigma_{OUT}$. In these cases, only dataset with a negative 
fluctuation of the ratio are included.

\subsection{Effect of the NXB spectrum renormalization}
\label{nxbreno}
As stated in Sect~\ref{results}, the NXB accounts for $\sim$80\% of the counts 
in the 2-8 
keV energy range for a blank sky observation.
It is therefore evident that even small errors in the renormalization of the 
NXB 
spectrum can yield large effects on the inferred CXB spectrum. The  
uncertainty on the renormalization factor, computed using 
standard error propagation, is of 0.8\% for both MOS1 and MOS2. 
This translates in an uncertainty of 2.1\% in the best fit photon index for 
the CXB spectrum, and of 1.9\% for the normalization.

\subsection{Check of the Vignetting curve calibration}
\label{vignetting}
The vignetting function of the X-ray telescopes is well calibrated (Lumb et 
al. 2002; Kirsch 2003); the main sources of uncertainty (currently under 
investigation by 
the calibration team) are a possible $\sim$1 arcmin offset of the optical 
axis from the nominal position and the azimuthal modulation induced by the 
Reflection Gratings Assemblies, installed on the light path of the X-ray 
telescopes having the MOS cameras at their primary foci. 
As a consequence, the absolute flux of a {\em point} source has an uncertainty 
of order $\leq$5\%.  
Since in this work we use events extracted from almost all 
of the MOS field of view, the uncertainty on the collected flux induced by 
the two quoted effects should be much lower, since possible errors in the 
vignetting corrections across the FOV should partially compensate. Following 
Lumb et al. (2002), we therefore assume 1.5\% as an estimate of the uncertainty 
on the effective areas.

\begin{figure}
\centering

  \resizebox{\hsize}{!}{\includegraphics[angle=-90]{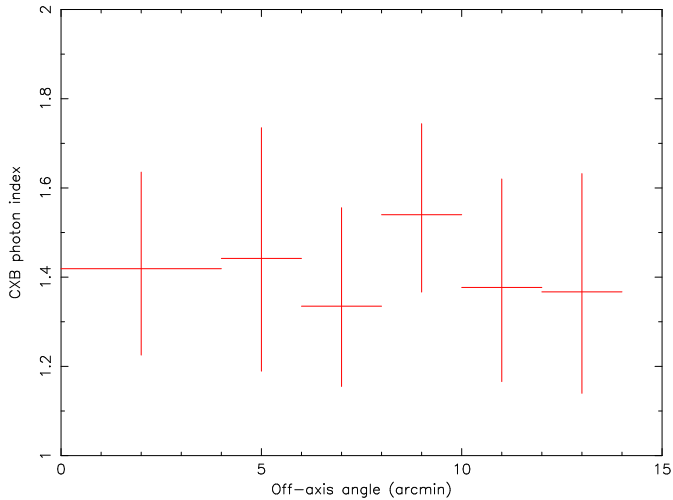}}
  \caption{ Best fit CXB photon index as a function of 
the off-axis angle. Each point represents an independent measure obtained 
from the analysis of a selected portion of the detector plane (see text). 
Errors (at 90\% confidence level for a single interesting parameter) include the 
uncertainty on the renormalization of the quiescent NXB 
spectrum. The MOS1 case is shown (MOS2 is very similar). The constant trend is 
a 
result of the correct vignetting function calibration.}
\label{slope_theta}
\end{figure}

\begin{figure}
\centering

  \resizebox{\hsize}{!}{\includegraphics[angle=-90]{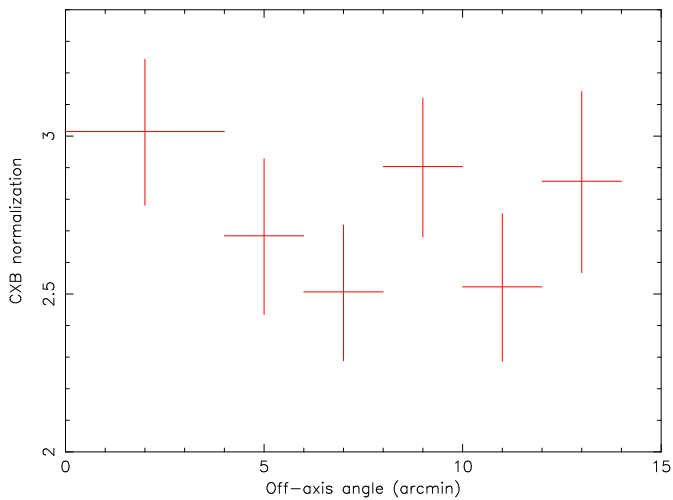}}
  \caption{ Same as Fig.~\ref{slope_theta}, the case of 
the CXB normalization is presented here. Units are photons cm$^{-2}$ s$^{-1}$ 
sr$^{-1}$ at 3 keV. The corrections for the light coming from out-of-field 
angles have not been applied. Errors (at 90\% confidence level for a single 
interesting 
parameter) include the uncertainty on the 
renormalization of the quiescent NXB spectrum. The variations are always 
smaller 
than the uncertainty, confirming the overall correctness of the vignetting curve
calibration.}
\label{norm_theta}
\end{figure}

As a  consistency check, we 
have divided the FOV of the instrument into six regions lying at 
increasing off-axis angles from the nominal optical axis position, namely a 
central circle of 4 arcmin radius and 5 circular annuli of 2 arcmin width. 
We repeated the whole analysis for each region separately, 
computing independent values for the best fit spectral parameters of the 
CXB. In Fig.~\ref{slope_theta} and ~\ref{norm_theta} we plotted the photon 
index and the normalization of 
the CXB as a function of the off-axis angle for the MOS1 camera; the MOS2 
case is very similar. As expected, no systematic effects can be 
seen, the observed fluctuations are smaller than the statistical 
uncertainties. 

 A study of the vignetting curve calibration using {\em slew} data is 
currently in progress. A preliminary analysis of $\sim$180 ksec of data
showed hints for deviations from the nominal behaviour in agreement with
the above quoted hypothesis of an offset of the aimpoint. The stray light
component (see next section), possibly centrally peaked, may also play
a role (D.Lumb, private communication). 

\subsection{Stray light}
\label{stray}
At the time of writing, no updates in the evaluation of the contribution  
to the collected flux from photons gathered from out-of-field angles have been 
published by the calibration team. We therefore assume the estimate by Lumb et 
al. (2002): an out-of-field scattered flux of order 7\% of the good focused 
in-field signal, with an associated systematic uncertainty of 2\%.  

 The main contribution to the stray light is due to sources within
0.4-1.4 degrees from the optical axis (Lumb et al. 2002). As a further check,
we have used the available data from past and current X-ray mission
to search for very bright sources lying in the quoted annular region
for each of the pointings.
We found only one relatively bright source at the $\sim3 \times 10^{-11}$
erg cm$^{-2}$ s$^{-1}$
level in the 2-8 keV energy range. A few (5) other sources were found, 
with fluxes of the order of a few $10^{-12}$ erg cm$^{-2}$ s$^{-1}$. 
According to the 
efficiency of the X-ray baffle quoted by Lumb et al. (2002), these should 
not represent a matter of concern for the evaluation of the CXB 
normalization.


\subsection{Effect of event PATTERN selection}
For a consistency check, we repeated the spectral analysis for the case of 
PATTERN 0 events only (i.e. single pixel events, see XMM-Newton Users' 
Handbook).
The results are reported in Table~\ref{tableresp0}, where the quoted errors 
are 
pure statistical uncertainties at the 90\% confidence level for a single 
interesting parameter. 

\begin{table*}[h]
\begin{center}
\begin{tabular}{cccc} \hline \hline
Instrument & Photon Index &   Normalization & $\chi^{2}_{\nu}$/d.o.f.  \\ 
\hline
MOS1   & 1.42$\pm$0.08 & 2.60$\pm$0.06 & 1.18/35  \\ \hline
MOS2      & 1.41$\pm$0.08 & 2.59$\pm$0.06  &  1.21/35   \\  \hline
MOS1+MOS2      & 1.42$\pm$0.06 & 2.596$\pm$0.043 & 1.16/72 \\ \hline \hline
\end{tabular}
\end{center}
\caption{\label{tableresp0} Results of the spectral analysis on the CXB for 
the 
MOS cameras. Only single pixel events (PATTERN 0) have been used. The 
normalization is expressed in photons cm$^{-2}$ s$^{-1}$ sr$^{-1}$ keV$^{-1}$ 
at 
3 keV (not corrected for the stray light contribution). The quoted 
uncertainties 
are purely statistical errors at the 90\% confidence level for a single 
interesting parameter.} 

\end{table*}

When the uncertainty associated to the renormalization of the quiescent NXB 
spectrum ($\sim$2\% for both the photon index and the normalization, see 
Sect.~\ref{nxbreno}) is taken into account, the results 
 from the PATTERN 0-12 data (Sect.~\ref{results}) and from PATTERN 0 events 
only 
are found to be in good agreement. We therefore assume any systematic effect 
associated with the event PATTERN selection to be negligible.

\subsection{Effect of the internal fluorescence line modeling}
 The fluorescence line component of the quiescent NXB varies strongly across 
the detector plane (see Sect.~\ref{spatial}). To account for this effect, 
we extract the source (sky fields) and background ({\em closed}) spectra 
from the same detector region. Residual differences in the line intensity are 
due to time variation of the fluorescence NXB (sky fields and {\em closed} 
observations are not symultaneous), possibly different with respect to the 
continuum NXB variation which is corrected by the {\em closed} spectrum 
renormalization. To solve the problem we have included multiple gaussian 
lines in the CXB spectral model (see Sect.~\ref{analysis}). 

To assess the possible effect of such correction on the CXB best fit 
parameters, we have repeated the spectral analysis removing the gaussian 
lines. In this case, the fit to the data is found to be somewhat worse 
($\chi^2_{\nu}=1.28$, 79 d.o.f.), owing to higher residuals in correspondence 
of the line energies. 
The best fit normalization is found to be in full agreement (within less 
than 0.1\%) with the value quoted in Sect.~\ref{results}, 
while the photon index is consistent within $\leq$2\% ($\leq~1 \sigma$
statistical uncertainty). 

We note that when the correction is applied, any systematics effect on 
the CXB spectral parameters should be smaller than the above quoted 
differences, and therefore negligible with respect to the pure statistical 
errors.

\subsection{Absolute flux calibration}
\label{absolute}
We can evaluate the uncertainty affecting absolute flux measurements with 
EPIC/MOS on the basis of the results of the cross-calibration with other 
satellites. Kirsch (2003) and Molendi \& Sembay (2003) reported a comparison 
of 
the spectral results on suitable calibration sources studied with different 
instruments. The agreement on the measured fluxes in the 1-10 keV range is 
very good, within a few \%, with a standard deviation of order 5\%. We assume 
5\% 
as a conservative estimate of the uncertainty on the absolute flux calibration 
of 
EPIC. 

\section{Discussion}
\label{discuss}
\begin{figure}
\centering

  \resizebox{\hsize}{!}{\includegraphics[angle=-90]{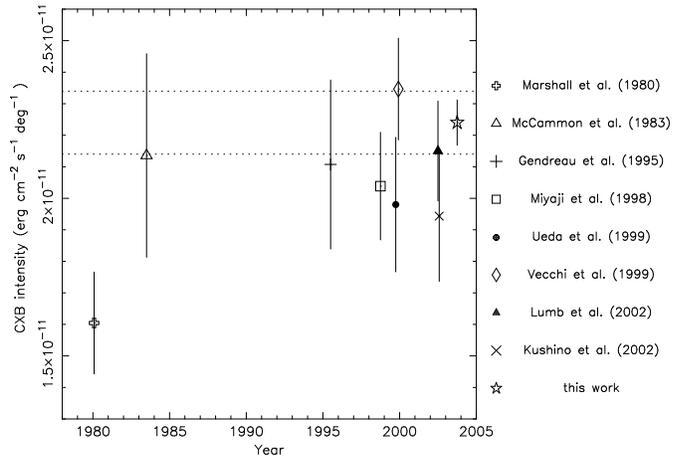}}
  \caption{CXB intensity measurements. The flux in the 2-10 keV band is 
represented 
 as a function of the epoch of the experiment. The plot is an update of Fig.3 of 
Moretti et al. (2003) including the results of the present work.  From left to 
right the CXB values are from Marshall et al. (1980) with HEAO-1 data; McCammon 
et al. 
(1983) with a rocket measurement; Gendreau et al. (1995) with ASCA SIS data; 
Miyaji et al. (1998) with ASCA GIS; 
Ueda et al. (1999) with ASCA GIS and SIS; Vecchi et al. (1999) with BeppoSAX 
LECS/MECS; Lumb et al. (2002) with XMM-Newton EPIC/MOS; Kushino et al. (2002) 
with 
ASCA GIS. Finally, the hollow star mark our own CXB measurement with the EPIC 
MOS 
cameras onboard XMM-Newton. All the uncertainties are at 1$\sigma$ level. 
The horizontal dotted lines mark the range of CXB intensity constrained by our 
measurement after including an extra 3.2\% uncertainty (1$\sigma$ confidence
 level, rescaled from the 5\% uncertainty quoted in the text, which corresponds
to 90\% confidence level) on the absolute flux 
calibration of the EPIC cameras (it is indeed a measure of the cross-calibration 
accuracy among different instruments). Such range represents the most robust
estimate of the true absolute sky surface brightness in the 2-8 keV band.
}
\label{cxbmeas}
\end{figure}

Our measurement of the intensity of the Cosmic X-ray Background has been 
plotted in Fig.~\ref{cxbmeas} (adapted from Moretti et al. 2003), together with 
previous determinations. 
 To be conservative, in order to identify the most probable range of
sky surface brightness constrained by our data, we have used an 
estimate of the accuracy of the cross-calibration in absolute flux among 
different 
recent instruments (see Sect.~\ref{absolute}) as a measure of the absolute flux 
calibration uncertainty of EPIC. 

Our results are in full agreement with the BeppoSAX LECS/MECS analysis of 
Vecchi et al. (1999) and with the study of Lumb et al. (2002) using XMM-Newton 
EPIC/MOS data. The ASCA GIS measure of Kushino et al. (2002) is marginally 
consistent with ours, while the HEAO-1 value (Marshall
 et al. 1980) is significantly lower.

As discussed by Barcons et al. (2000), part of the dispersion of the different 
measurement can be ascribed to cosmic variance. The above authors estimated 
that differences up to $\sim$10\% in the CXB intensity are expected when the 
solid 
angle is of 
order 1 square degree. They concluded therefore that systematic errors and 
cross-calibration differences among different instruments must be present.

The HEAO-1 measurement, performed on$\sim10^4$ square degrees, is the most 
robust as for solid angle coverage. However, all of the subsequent 
measurements 
yielded systematically higher values, casting at least some doubts on the 
absolute flux calibration of the HEAO-1 instruments. We note that a recent 
reanalysis of the HEAO-1 data (Gruber et al. 1999), as
pointed out by Gilli (2003), showed differences of order 
10\% among different detectors in the overlapping energy ranges.

Two new measurements have been published since the analysis of Barcons 
et al. (2000). Lumb et al. (2002) used a combination of 8 XMM-Newton EPIC/MOS 
pointings covering $\sim1.2$ square degrees. Their approach for the 
instrumental 
background subtraction (based on a model of the {\em out FOV} spectrum) is 
markedly different from ours and therefore their determination is to be 
considered largely independent from the measurement presented in this work.
The ASCA/GIS study of Kushino et al. (2002) was performed over a large solid 
angle ($\sim$50 square degrees) to minimize the effects of cosmic variance. 
However, the large stray light component ($\sim$40\% of the collected flux) 
affecting ASCA represents a severe limit to the absolute flux measurement of 
the CXB, which they estimated to be uncertain at the $\sim$10\% level.

In our analysis, we used a compilation of sky pointings covering $\sim5.5$ 
square degrees. The effects of cosmic variance (roughly scaling as $\Omega^{-
1/2}$) on our measure of the CXB intensity should be rather small, being of 
the 
same order of the overall quoted uncertainty ($\sim4$\%). Our measurement, 
performed with the well-calibrated EPIC/MOS instrument, relies on a very 
robust 
characterization and subtraction of the instrumental background; possible 
sources of systematics were carefully studied and their impact on the 
determination of the CXB spectrum was evaluated.
In conclusion, these considerations, coupled to the excellent agreement of our 
findings with the results of two of the most recent investigations, lead us to 
believe that the measurements of the CXB intensity have finally converged to a 
well constrained value, significantly higher than the former result from 
HEAO-1 data, assumed more than 20 years ago as a reference.

It is very interesting to compare our new measurement of the CXB intensity 
with 
the source number counts derived from the recent deep X-ray observations. 
Moretti et al. (2003) have built a Log$N$/Log$S$ function using a very large 
source 
compilation, including the results from six different surveys, both 
pencil-beam 
and wide-field, performed with ROSAT, Chandra and XMM-Newton. We assume here 
their results for the integrated flux of the 2-10 keV source counts down to 
the 
sensitivity limit of the 1 Ms Chandra Deep Fields ($\sim1.7\times10^{-16}$ erg 
cm$^{-2}$ s$^{-1}$). Our measurement of the CXB intensity, 
F$_{CXB}$=(2.24$\pm$0.10)$\times10^{-11}$ erg cm$^{-2}$ s$^{-1}$ deg$^{-1}$
 ($1\sigma$ error, including the absolute flux uncertainty), 
implies that 80$^{+7}_{-6}$\% of the cosmic X-ray background has been 
resolved into discrete sources in the 2-10 keV band. Even extrapolating the 
Log$N$/Log$S$ down to fluxes of $\sim10^{-17}$ erg cm$^{-2}$ s$^{-1}$ (a 
factor 
of 10 below the detection limit of the deep surveys) the integrated flux would 
rise only to $\sim$84\% of the CXB (due to the slow growth of the faint branch 
of the Log$N$/Log$S$), not consistent with its total value. A new class of 
faint 
sources (possibly heavily absorbed AGNs, or star-forming galaxies, see Moretti 
et al. 2003 and references therein) could emerge at fluxes of a few  $10^{-16}$ 
erg cm$^{-2}$ s$^{-1}$, steepening the Log$N$/Log$S$ and accounting for the 
remaining part of unresolved CXB. Some contribution to the CXB could also be 
due 
to truly diffuse emission.

The analysis of the deepest 2 Ms Chandra Deep Field North and of the XMM deep 
pointing to the Lockman Hole and of their associated multiwavelength follow-up 
campaign, will shed light on the issue.
On the basis of a well constrained value for the CXB intensity, the new 
observations will possibly allow for an ultimate understanding of the CXB 
nature.

\appendix

\section{An independent measure of the quiescent NXB spectrum}
\label{qnxb}
The quiescent NXB component due to the interaction of high-energy particles 
with the detectors and the other structures of the telescopes/cameras 
bodies can be studied in two ways: (i) considering the {\em out FOV} regions 
of 
the detectors and (ii) using the {\em closed} observations. In both cases, the 
collected signal is purely cosmic-ray induced, with no contributions from 
X-ray photons or soft protons. 

To the aim of a measure of the CXB spectrum, {\em out FOV} regions offer the 
advantage of a NXB measurement simultaneous with the CXB observation. A 
disadvantage is represented by possible spatial variations of the NXB 
spectrum across the plane of the detectors: in such case, the {\em out FOV} 
spectrum would not be an adequate representation of the NXB present in an 
observation of the sky. Closed observations allow for the extraction of 
the NXB spectrum from the same detector region where the CXB is collected, 
but the two measurements are not simultaneous. Time variations of the NXB 
spectrum may render {\em closed} observations useless for our study. 
These problems will be addressed in the following sections.

\subsection{Spectral characterization of the quiescent NXB}

 The spectrum extracted from the merged {\em closed} dataset, selecting the   
{\em in FOV} region, is presented in Fig.~\ref{qnxbspectrum}. The total exposure 
time 
is of 
$\sim$430 ksec. In the same plot we show the spectra extracted from the {\em 
out 
FOV} region for the sky fields merged dataset (total exposure $\sim$1.15 Ms) 
and 
the {\em closed} dataset.
The 2-12 keV continuum is well represented by a flat power law not convolved 
with the telescope collecting areas and CCD quantum efficiencies, with a 
photon 
index of $\sim$0.2.
Several bright emission lines are seen, due to fluorescence emission from 
materials of the telescope and 
the camera bodies. The strongest are the Al-K line from the shielding 
of the cameras and the Si-K line from the back substrate of the CCD chips (its 
intensity is much smaller in the {\em out FOV} spectrum); 
additional lines from Cr, Mn, Fe, Zn, Au are also visible. For further details 
see Lumb et al. (2002).

\subsection{Spatial distribution of the quiescent NXB}
\label{spatial}
It is known that the quiescent NXB changes slightly across the field of view 
(e.g. Lumb et al. 2002). The strongest spatial variations are seen in 
correspondence with the internal fluorescence emission lines. Lumb et 
al. (2002) 
showed the peculiar, non-uniform distribution of the Al-K (strongly {\em 
suppressed} near the edges of CCD\#1) and of the Si-K (strongly {\em brighter} 
near the edges of CCD\#1) emission (see their Figures 2 and 3), due to 
shadowing 
effects as a consequence of the non-planar disposition of the CCD chips.

\begin{figure}
\centering

  \resizebox{\hsize}{!}{\includegraphics[angle=-90]{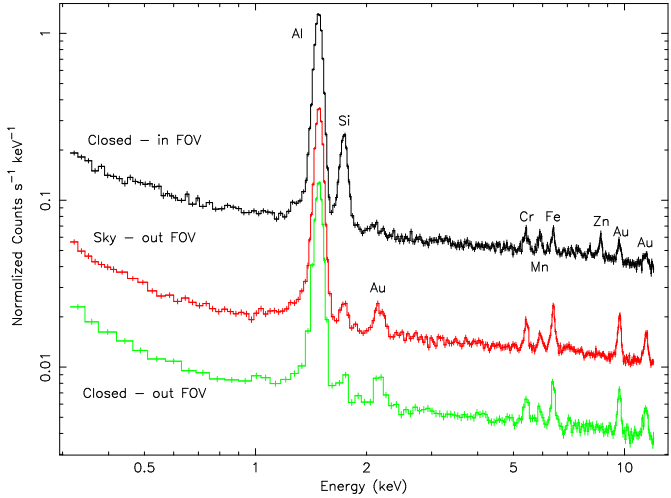}}
    \caption{ Quiescent NXB spectrum for the MOS1  
camera. From top to bottom: data from {\em closed} observations (black), {\em 
in 
FOV} region; from the sky merged dataset (red), {\em out FOV} region; from 
{\em 
closed} obs.(green), {\em out FOV} region. 
The {\em closed out FOV} spectrum has been rescaled by a factor of 3 to ease 
its 
visibility; no renormalization was performed on the other spectra.
Total exposure time 
is of $\sim$430 ksec for the {\em closed} spectra, $\sim$ 1.15 Ms for the sky 
data. The labels identify the strongest internal fluorescence lines.
Significant differences in the fluorescence lines intensities are evident 
between the {\em in FOV} and the {\em out FOV} spectra.}
\label{qnxbspectrum}
\end{figure}

An inspection of Fig.~\ref{qnxbspectrum} immediately shows the different 
relative strength of the fluorescence lines in the quiescent {\em in FOV} 
and {\em out FOV} NXB. Apart from the above mentioned Al-K and Si-K cases, it is 
evident that the Au emission (most probably associated with the  gold coated 
readout
flexible circuits) 
is much stronger in the {\em out FOV} spectrum. In particular, the 
prominent structure at $\sim$2.15 keV in the {\em out FOV} spectrum is 
virtually 
absent {\em in FOV}.

To investigate the possible presence of spatial variations in the continuum, 
we 
have studied the {\em closed} NXB on each CCD detector separately. The spectra 
were fitted in the 2-12 keV range with a model consisting of a power law with 
the addition of multiple gaussians to reproduce the fluorescence lines, not 
convolved with the telescope areas and the CCD quantum efficiencies. We found, 
as expected, strong variations in the line intensities (their energies and 
width 
are found to be stable); marginally significant variations in the photon index 
were also seen.

The observed variations of the quiescent NXB across the FOV strongly suggest 
that it is better to extract the NXB and the source spectra from the same 
region 
of the detector. This is particularly important in the case of the faint CXB, 
whose signal is lower than the NXB over the whole 2-8 keV energy range.

\begin{figure}
\centering

  \resizebox{\hsize}{!}{\includegraphics[angle=-90]{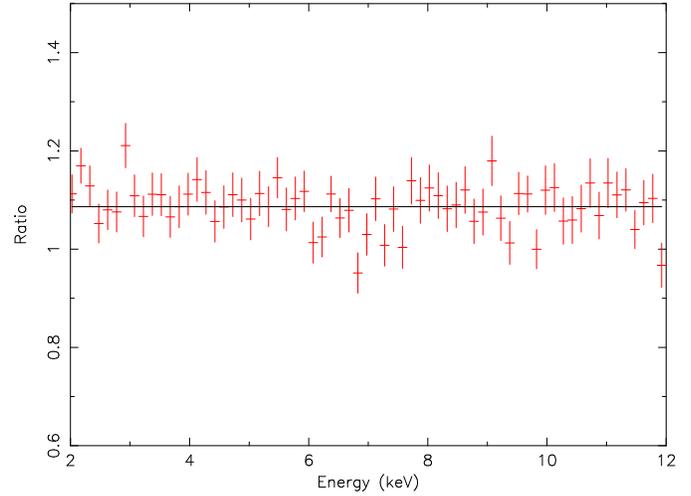}}
\caption{ Ratio between the spectra extracted from the {\em out 
FOV} region for the sky merged dataset and for the {\em closed} dataset. Data 
are from 
the MOS1 camera; the MOS2 case is very similar. A 
linear fit to the range 2-12 keV is fully consistent with a constant.}
\label{ratio_out}
\end{figure}

To assess whether the {\em closed} observations are indeed representative of 
the 
actual quiescent NXB affecting a typical sky field observation, we compared 
the 
{\em closed} spectrum extracted from the {\em out FOV} region with the 
corresponding {\em out FOV} spectrum from the merged sky dataset. Apart from a 
multiplicative factor required to match the intensities (due to the 
non-simultaneity of the measurements), we did not find significant differences 
either in the continuum shape (the best fit photon indexes were almost 
indistinguishable) or in the line components. In Fig.~\ref{ratio_out} we 
show the ratio between the two spectra. The results of a linear fit to the 
ratio 
in the range 2-12 keV are fully consistent with a constant.

\subsection{Temporal behaviour of the quiescent NXB}
\label{timenxb}
\begin{figure}
\centering

  \resizebox{\hsize}{!}{\includegraphics[angle=0]{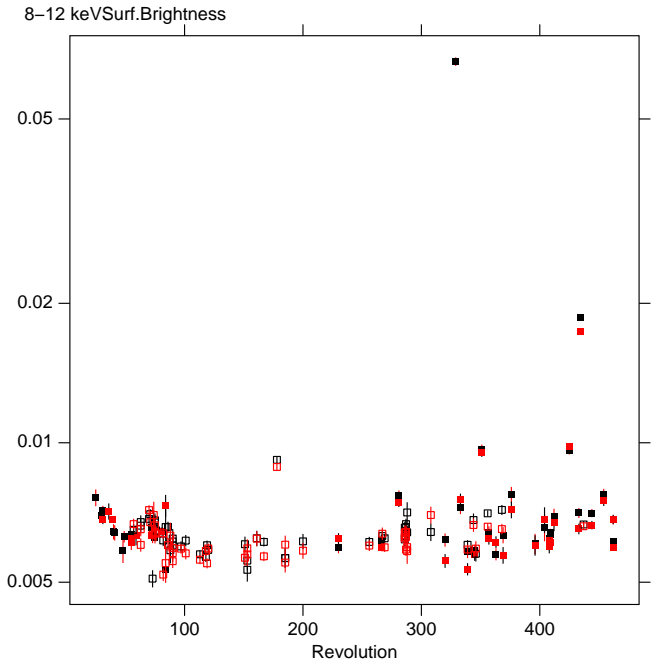}}
  \caption{Secular variation of the surface brightness 
  (units of counts cm$^{-2}$ s$^{-1}$) in 
the 8-12 keV 
range for MOS1 (red points) and MOS2 (black points): {\em out FOV} 
regions in sky fields observations (empty squares) and  {\em in FOV} region 
in  
{\em closed} observation (filled squares).  A smooth trend having a broad 
minimum around revolution $\sim$150 is seen; a high scattering, due to 
different particle radiation conditions, is also evident.}
\label{secularopen}
\end{figure}

Non-simultaneous measurements of the NXB can be used to study 
the CXB only if the NXB spectrum is stable in time.
As a first step, we studied the secular variation of the NXB surface 
brightness in different energy ranges. As a typical example, we show in 
Fig.~\ref{secularopen} the light curves in the 
8-12 keV band for the {\em out FOV} regions of the sky observations and for 
the 
{\em closed} observations (the {\em in FOV} region was selected in this case).

The main results are:

\begin{itemize}
\item the large majority of the observations follow a smooth secular 
trend: the surface brightness decreases by 30\% from revolution 50 to 
revolution 100, reaches a minimum around revolution 150 and thereafter 
slowly increases.

\item the secular trend and the absolute surface brightness values for the 
normally behaving cases are almost equal for the sky field data {\em out FOV} 
and the 
{\em closed} observations.

\item The points deviating from the general 
trend correspond to observations performed under high radiation 
conditions. For instance, the {\em closed} observation of revolution 329 was 
taken 
during an intense solar flare.

\item during high radiation periods, the intensity of the NXB can be more than 
an order of magnitude higher than average.

\item the scatter around the good behaviour is of order 15\% for the 
sky observations, while it is much higher for the {\em closed} observations, 
since 
the latter are often performed at the beginning/end of the visibility window 
of 
the orbit, closer 
to the perigee, where the flux of particles (from the radiation belts) is 
generally more intense than in the higher part of the satellite orbit.

\end{itemize}

\begin{figure}
\centering

  \resizebox{\hsize}{!}{\includegraphics[angle=0]{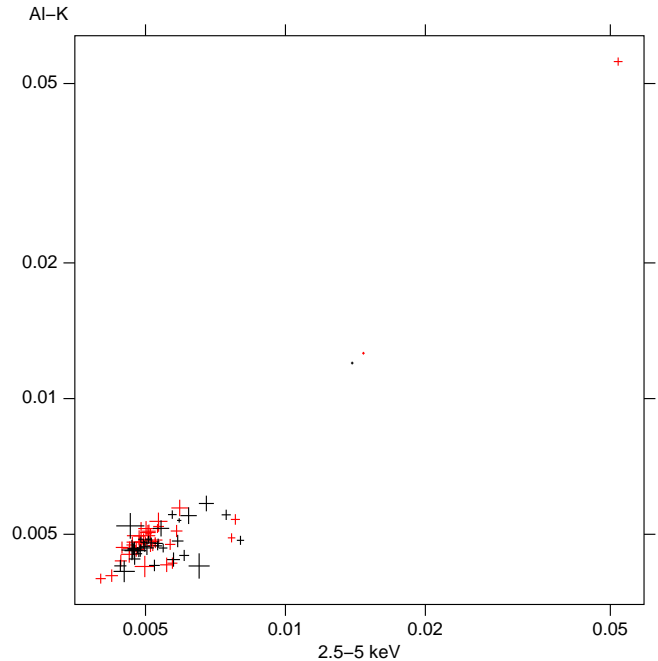}}
  \caption{The correlation between the surface brightness  (units of counts 
cm$^{-2}$ 
s$^{-1}$) 
in the 1.4-1.6 keV range (dominated by the internal Al-K fluorescence line - 
see Fig.~\ref{qnxbspectrum}) and the surface brightness in the range 2.5-5 
keV (continuum). Data are for {\em closed} observations; MOS1 red and MOS2 
black. 
A rather high scatter is evident.}
\label{al_vs_cont} 
\end{figure}

\begin{figure}
\centering

  \resizebox{\hsize}{!}{\includegraphics[angle=0]{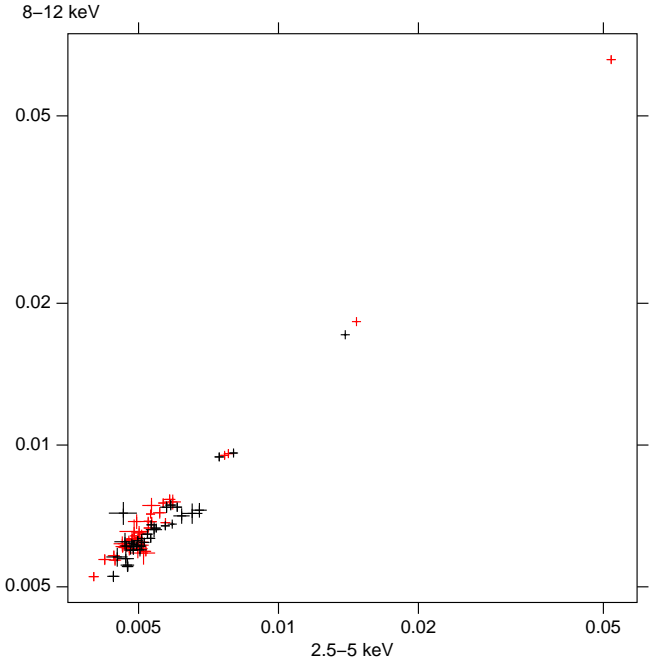}}
  \caption{ Same as Fig.~\ref{al_vs_cont}; the surface 
brightness  (units of counts cm$^{-2}$ s$^{-1}$) in the energy ranges (2.5-5 
keV) and 
(8-12 keV) show a better 
correlation.}
\label{cont1_vs_cont2}
\end{figure}

The light curves in other energy bands are very similar and are not 
reported here. What is important to the aim of the present study (as 
well as for the study of faint diffuse sources) is to assess the stability 
of the quiescent NXB spectral shape. In Fig.~\ref{al_vs_cont} we plot the 
surface 
brightness in 
the 
band 1.4-1.6 keV (dominated by the Al-K internal fluorescence line) versus 
the 
2.5-5 keV band (continuum emission only) for the {\em closed} observations. It 
is evident that, although a definite correlation is present, the scatter of 
the points is rather high. A very similar effect is seen comparing the 
0.5-1.2 keV band with the 2.5-5 keV; on the contrary, the correlation 
between the 2.5-5 keV and the 8-12 keV bands is much better (see 
Fig.~\ref{cont1_vs_cont2}).

This suggests that the spectral shape may change in time, the fluorescence 
lines 
having a time behaviour quite different from the high energy continuum. A 
similar result 
was reported for the pn camera by Katayama et al. (2002). A poor correlation is 
also seen between the low energy and the high energy parts of the 
continuum, while the 2-12 keV band presents a rather coherent behaviour. 
These results have important implications for the study of faint diffuse 
sources when an independent measurement of the background spectrum is 
needed: a simple renormalization of the quiescent NXB using the high energy 
range (e.g. 8-12 keV) count rate could lead to 
systematic errors in both the continuum (low energy wrt. high energy) and 
the 
lines.

\begin{figure}
\centering

  \resizebox{\hsize}{!}{\includegraphics[angle=-90]{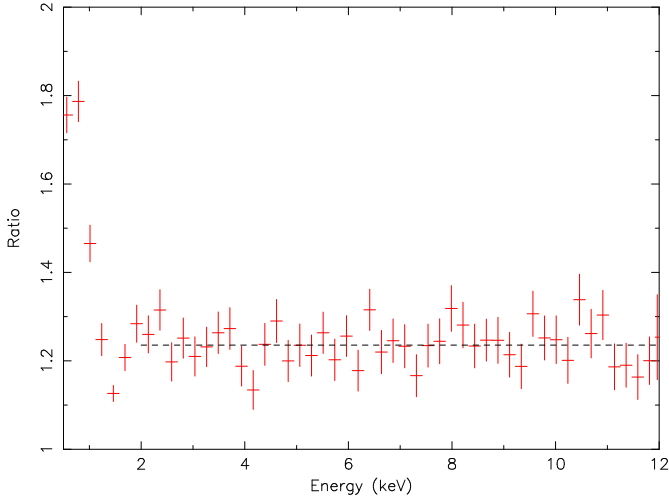}}
  \caption{ Ratio of the spectra from the ``high 
intensity'' and ``low intensity'' samples of {\em closed} observations (see 
text) for the MOS1 camera. The MOS2 case is very similar.
Although deviations are seen in correspondence of the fluorescence lines, 
the overall shape in the 2-12 keV range is clearly constant: the shape of 
the quiescent NXB spectrum is stable in spite of high variations in the flux 
of high energy particles.}
\label{hi_vs_low}
\end{figure}

We investigated in more detail the dependence of the NXB spectral shape on the 
intensity of the high-energy particle flux. We divided the sample of the {\em 
closed} observations in two 
parts, a ``high intensity'' dataset and a ``low intensity'' dataset 
(respectively, above and below a 8-12 keV surface brightness of 0.007 cts 
cm$^{-2}$ s$^{-1}$, see Fig.~\ref{secularopen}). We 
then extracted the spectra from the {\em in FOV} region. We show in 
Fig.~\ref{hi_vs_low} 
the ratio of the two spectra. As expected from the previous analysis, 
differences are observed in the low energy part of the continuum and in 
correspondence of the brightest fluorescence lines; however, the most 
interesting result is that the ratio above 2 keV is constant. We tried a 
linear fit to the data in the 2-12 keV range; the slope was found to be 
statistically null.

This demonstrates that the NXB continuum spectral shape in this energy 
range does not depend on the intensity of the impinging flux of cosmic rays; 
no significant changes are 
seen even in presence of very large variations in intensity. It is 
therefore possible to use non-simultaneous measures of the quiescent NXB 
spectrum for the study of the 2-8 keV CXB. The only caveat is to compute a 
renormalization factor in order to match the intensity of the NXB 
component present in the two dataset.

\subsection{Conclusions}
The results of the study of the quiescent NXB can be summarized as 
follows.

\begin{itemize}

\item The ratio between the {\em closed} and the sky field ({\em out FOV}) 
spectra shows that {\em closed} observations 
give a correct representation of the quiescent NXB present in a typical 
observation of the sky.

\item The spatial analysis revealed strong variations in the fluorescence line 
component across the detector plane, as well as hints for variations in the 
continuum. A NXB spectrum 
extracted from the same detector region where the CXB is collected would 
therefore be preferable.

\item The temporal analysis demonstrated that variations up 
to a factor of 10 in intensity may occur during high radiation periods. The 
spectral shape may change in time; in particular, the low energy (0.5-1.2 keV) 
and high energy ($>$2 keV) continuum variations are poorly correlated; the 
same 
is true for continuum and fluorescence line component. In any case, the 
spectral 
shape above 2 keV is very stable, at least for the continuum.

\end{itemize}

In conclusion, to the aim of the present work, the use of the {\em closed} 
observations is  more appropriate 
to get the indepenent measurement of the quiescent NXB required to study 
the CXB.


\section{Any residual soft proton contamination ?}
\label{aqnxb}

A very important step in order to get a robust measure of the faint CXB 
spectrum is to investigate the presence of a residual component of soft 
proton NXB in the GTI-filtered event lists, as a consequence of 
a steady flux of particles impinging the 
detectors and/or of other peculiar conditions of low-energy particle 
irradiation.
A soft proton contamination could indeed affect the study of faint 
diffuse sources; so far, no systematic investigations of this problem 
were published (but see De Luca \& Molendi 2002 for a preliminary 
report). 
 
In the case of the ACIS instrument onboard Chandra, rather similar to 
XMM-EPIC in both 
the orbit and the detector technology, Markevitch et al. (2003) reported 
that the spectrum of the dark Moon is consistent with the spectrum 
collected when the CCDs are in the stowed position (outside the field of 
view) and therefore the steady flux of low energy particles can be as 
low as zero.
In any case, the low statistic in the dark Moon spectra, coupled to the 
small number of observations, prevented the authors from drawing definitive 
conclusions. However, Markevitch et al. (2003) showed that ``long flares'', 
having a very low time variability along an observation, may occasionally 
occur, affecting the spectral study of faint diffuse sources.
The higher collecting area suggests a priori 
that in similar conditions the flow of gathered particles could be higher (and 
therefore its 
impact on 
science more important) for XMM/EPIC than for 
Chandra/ACIS.

In the next sections we will present a simple diagnostic to reveal and 
quantify the residual soft proton component affecting an observation; we 
will also study its impact on the measure of the CXB spectrum. 

\subsection{A diagnostic to reveal soft proton contamination}
\label{existenceaqb}
The basic idea to identify the residual low energy particle  background 
is to look for its signature in terms of a non-uniform distribution of 
the count rates across the plane of the detectors, due to the focusing 
of the particles by the telescope optics. We studied therefore the 
surface brightness in the regions inside and 
outside the field of view: any residual soft proton component  
 should be visible only {\em in FOV}. We selected the high 
energy band (8-12 keV), excluding the central CCD in order to minimize 
 the presence of genuine cosmic X-rays in the count rate {\em in FOV}.

\begin{figure}
\centering

  \resizebox{\hsize}{!}{\includegraphics[angle=0]{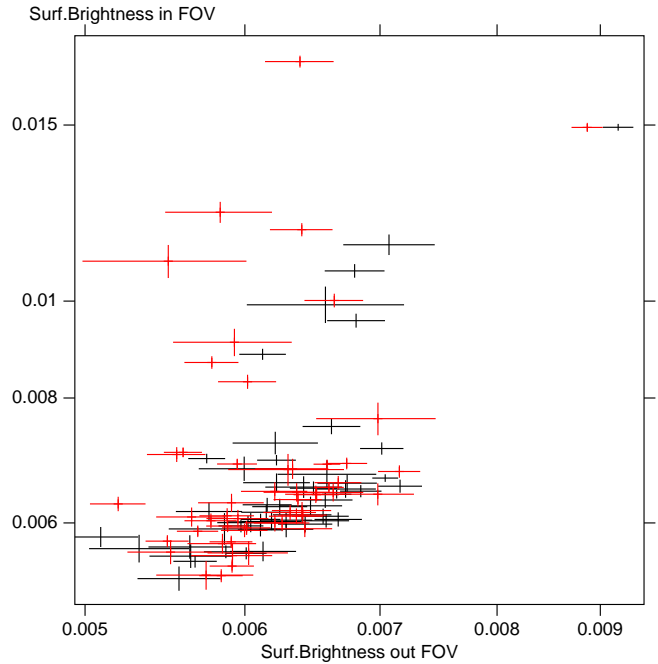}}
  \caption{ The correlation between the 8-12 keV surface 
brightness (counts cm$^{-2}$ s$^{-1}$) {\em in FOV} and {out FOV} is shown for 
the sky 
fields 
observations. Red and black points represent MOS1 and MOS2 observations, 
respectively. A high scatter is clearly evident.}
\label{in_vs_out}
\end{figure}

\begin{figure}
\centering

  \resizebox{\hsize}{!}{\includegraphics[angle=0]{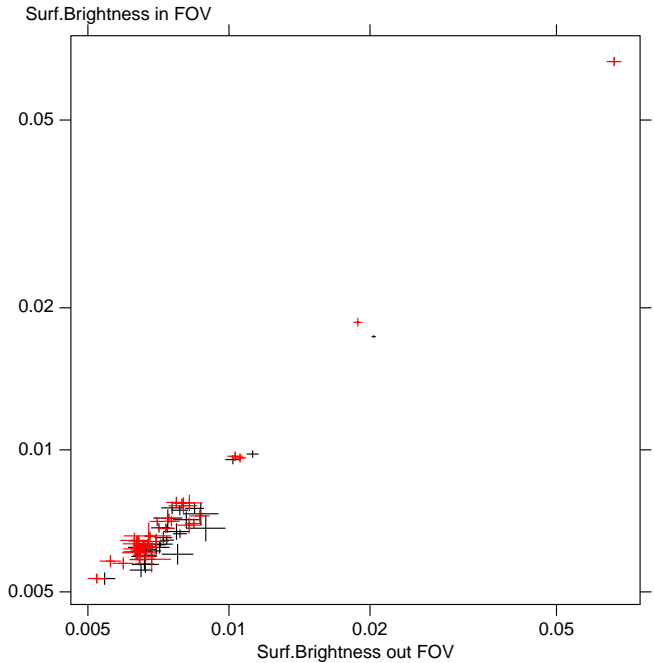}}
  \caption{ Same of Fig.~\ref{in_vs_out} for the case of 
the {\em closed} observations. The correlation is much better.}
\label{in_vs_out_cl}
\end{figure}

As a first step, we plotted the surface brightness {\em in FOV} 
($\Sigma_{IN}$) 
versus the surface brightness 
{\em out FOV} ($\Sigma_{OUT}$). This is shown in Fig.~\ref{in_vs_out}. A 
clear correlation is 
present; however, a rather high scatter of the points is evident. The 
same plot obtained for the case of the {\em closed} observations shows a 
much better correlation (see Fig.~\ref{in_vs_out_cl}).

\begin{figure}
\centering

  \resizebox{\hsize}{!}{\includegraphics[angle=0]{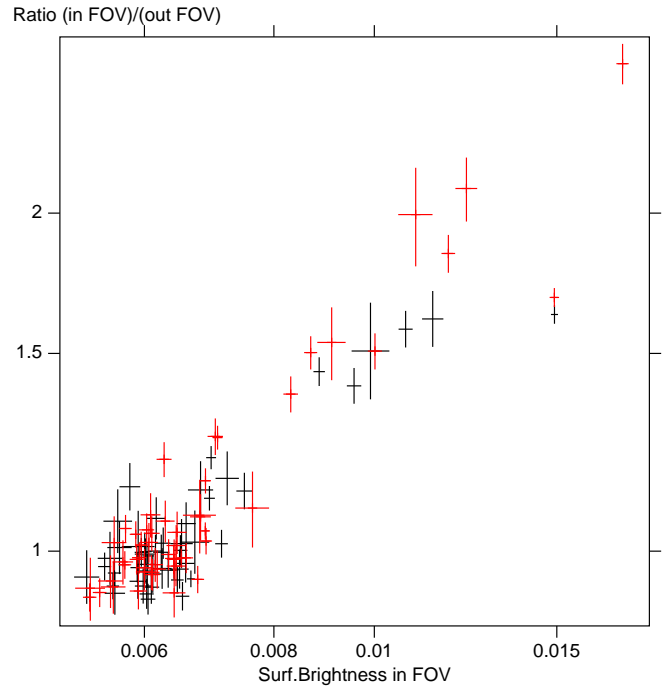}}
  \caption{ The ratio of the 8-12 keV surface brightness 
({\em in FOV})/({\em out FOV}) is plotted as a function of the surface 
brightness {\em in FOV} (counts cm$^{-2}$ s$^{-1}$). Data are from sky fields 
observations, red points 
correspond to MOS1 and black points to MOS2. A definite correlation is 
absolutely evident.}
\label{ratio_vs_in}
\end{figure}

\begin{figure}
\centering

  \resizebox{\hsize}{!}{\includegraphics[angle=0]{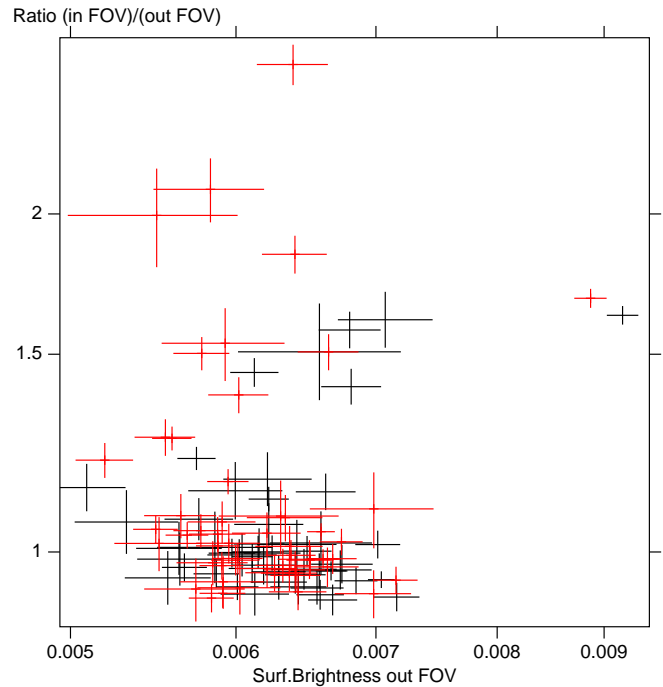}}
  \caption{ The ratio $\Sigma_{IN}/\Sigma_{OUT}$ (8-12 
keV)  is plotted here as a function of $\Sigma_{OUT}$ (counts cm$^{-2}$ 
s$^{-1}$). No 
correlation can be 
seen in this case, markedly different from the result presented in 
Fig.~\ref{ratio_vs_in}.  }
\label{ratio_vs_out}
\end{figure}

As a second step, we studied whether the scatter in Fig. 
~\ref{in_vs_out} can be ascribed to an anomalous flux {\em in FOV}.
This is absolutely clear from Fig.~\ref{ratio_vs_in}, where we have 
plotted the ratio of the surface brightness $\Sigma_{IN}/\Sigma_{OUT}$ 
versus the surface brightness $\Sigma_{IN}$. Such a clear correlation is not 
seen at all in the plot 
of $\Sigma_{IN}/\Sigma_{OUT}$ vs. $\Sigma_{OUT}$ presented in 
Fig.~\ref{ratio_vs_out}.

It is thus evident that in some observation the NXB surface brightness
inside the field of view is higher than outside. From Fig. 
~\ref{ratio_vs_in}, we can see that the most pathological cases can show 
a quiescent NXB intensity up to 300\% higher than average. We 
note that 
such anomalous count rates are to be entirely ascribed to an enhanced 
NXB component, since the signal expected from the CXB in the selected 
energy range and off-axis angle is of order of 
$\sim$0.5\% of the overall count rate, and no bright sources are 
present in our selected fields. 

\begin{figure}
\centering
  \resizebox{\hsize}{!}{\includegraphics[angle=-90]{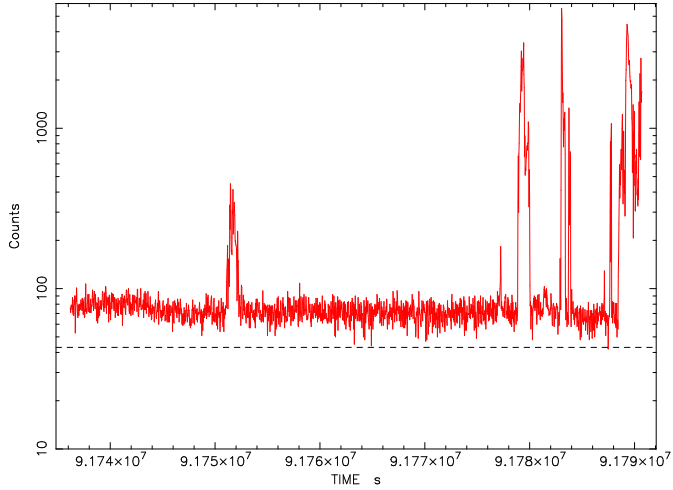}}
  \caption{ The light curve of this observation looks 
``typical'' at first glance, with intense soft proton flares superposed on a 
quiescent count rate. However it is found that the quiescent count rate is 
much 
higher than the value expected for a typical blank sky field (marked by the 
dashed line) and shows a slow modulation. The whole observation is affected by 
a 
``long flare'' of low-energy particles; the analysis of the 8-12 keV surface 
brightness after GTI screening clearly indicates an anomalous ratio 
$\Sigma_{IN}/\Sigma_{OUT}$.}
\label{lc_smooth}
\end{figure}

\begin{figure}
\centering
\resizebox{\hsize}{!}{\includegraphics[angle=-90]{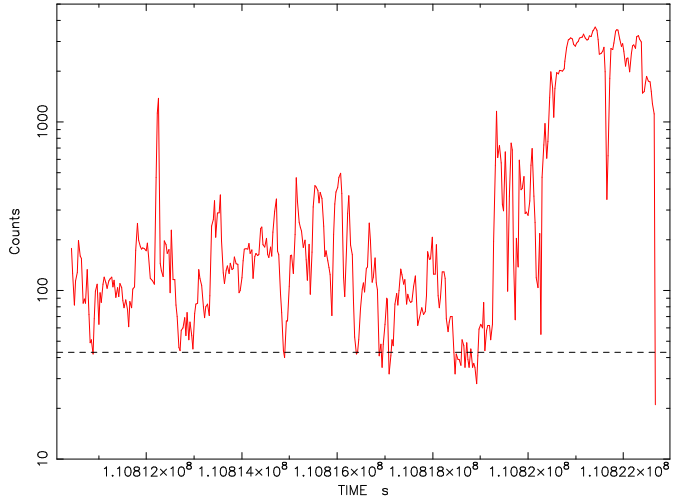}}
\caption{ Light curve of an observation badly affected by 
soft proton flares during all of its duration. The expected count rate for a 
typical sky field is marked by the dashed line. The automated GTI filtering 
algorithm used in our study cannot properly work in such conditions (see 
text).}
\label{lc_complicated}
\end{figure}

We inspected the light curves of the most deviating observations. Two 
phenomenologies, as expected, were observed. First, a very smooth light 
curve, with an average ``quiescent'' count rate significantly higher than 
expected for a typical 
blank sky field, and slow time variability along the observation duration 
(see Fig.~\ref{lc_smooth}). These could be the cases of ``long flares'' of 
soft 
protons, 
affecting the whole observation and showing little or no time variability, 
similar to what occasionally observed by Markevitch et al. (2003) in the 
Chandra case. Second, conversely, a highly variable light curve with 
almost no quiescent time intervals (Fig.~\ref{lc_complicated}). In both cases, 
as stated in 
Sect.~\ref{check}, our automated GTI filtering algorithm cannot be 
efficient since all of the observing time is affected by a flux of low 
energy particles.

\begin{figure}
\centering

  \resizebox{\hsize}{!}{\includegraphics[angle=-90]{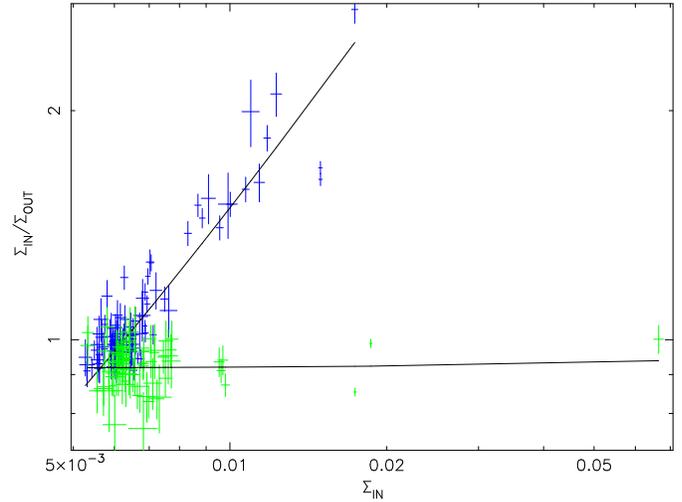}}
  \caption{The ratio of the 8-12 keV surface brightness 
$\Sigma_{IN}/\Sigma_{OUT}$ is plotted as a function of $\Sigma_{IN}$ (counts 
cm$^{-2}$ 
s$^{-1}$) for the 
sky fields (blue points) and for the {\em closed} observations (green points). 
A 
linear fit to the two distributions is overplotted. It is evident that the 
sky fields observations show higher values of $\Sigma_{IN}/\Sigma_{OUT}$ 
even in the less contaminated cases.}
\label{ratio_fit} 
\end{figure}

In Fig.~\ref{ratio_fit} the same plot as in Fig. 
~\ref{ratio_vs_in} is presented; data from {\em closed} observations have been 
overplotted. It is evident that for {\em closed} observations the ratio of 
the surface brightness $\Sigma_{IN}/\Sigma_{OUT}$ remains constant under all 
of the 
observed low energy particle irradiation conditions. The results of a 
linear fit to the two 
distributions are also interesting: the ratio  $\Sigma_{IN}/\Sigma_{OUT}$
is found to be lower for the {\em closed} observations. 
This is true also for the sample of sky fields observations which are less 
affected by the residual soft proton component. In correspondence of the 
average 
value of the surface brightness {\em in FOV} for these observations 
($\Sigma_{IN}\sim0.0065$ cts cm$^{-2}$ s$^{-1}$ in 8-12 keV), the ratio 
$\Sigma_{IN}/\Sigma_{OUT}$ for the sky fields is higher by $\sim10$\%. 
Although 
this difference is seen at a low level of significance ($<$3$\sigma$), this 
may 
be a hint for the presence of an irreducible flow of low energy particles 
reaching the detectors during all  science observations.

\subsection{Impact on science of the residual soft proton component}
\label{impact}
In the previous section we showed that a residual component of 
soft proton background is sometimes (possibly always) present within the field 
of view. It is very important to assess whether such NXB 
component can affect the scientific analysis 
of faint diffuse sources, and, if this is the case, 
to define a recipe to recover reasonable results from 
contaminated observations.

\begin{figure}
\centering

  \resizebox{\hsize}{!}{\includegraphics[angle=-90]{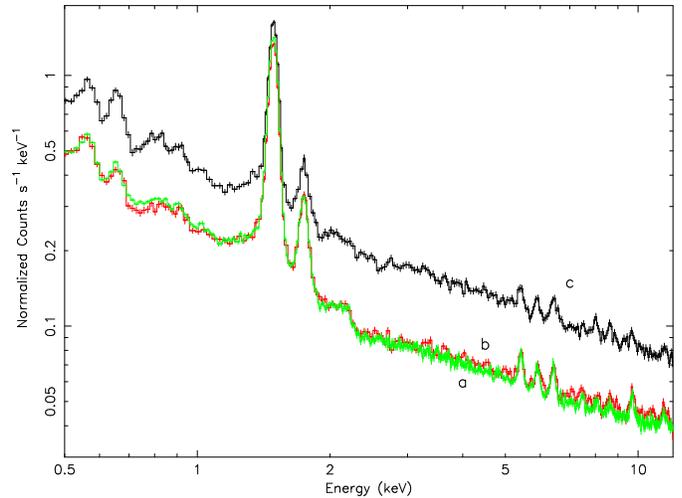}}
  \caption{ Spectra extracted from the sky 
fields observations having different values of the ratio $R$ of surface 
brightness $\Sigma_{IN}/\Sigma_{OUT}$: R$<$1.05 (green - dataset ``a''), 
1.05$<$R$<$1.30 (red - dataset ``b''), R$>$1.30 (black - dataset ``c'').}
\label{spectra_differentnxb}
\end{figure}

\begin{figure}
\centering

  \resizebox{\hsize}{!}{\includegraphics[angle=-90]{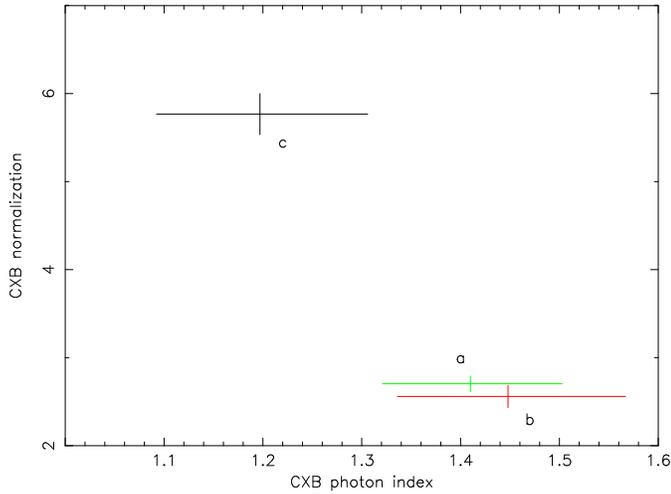}}
  \caption{Best fit CXB parameters 
computed from the three independent dataset corresponding to different 
values of $\Sigma_{IN}/\Sigma_{OUT}$ (see text and 
Fig.~\ref{spectra_differentnxb} for the color code. The normalization is in 
photons cm$^{-2}$ s$^{-1}$ sr$^{-1}$ keV$^{-1}$ at 3 keV. The stray light 
contribution have not been subtracted. Errors (at 90\% confidence level for a 
single 
interesting parameter) include the uncertainty on the 
renormalization of the quiescent NXB spectrum. The normalization and 
the photon index computed for the dataset (c) are uncompatible with the values 
for dataset (a) and (b), which are found, conversely, to be in good 
agreement.}
\label{cont_spectra_differentnxb} 
\end{figure}

To this aim we divided our sample of 
observations in different independent subsets, corresponding to different 
levels of contamination, the ratio $R$ of surface brightnesses 
$\Sigma_{IN}/\Sigma_{OUT}$ being a measure of the residual SP NXB level. 
We selected three ranges of $R$: $<$1.05 (negligible contamination - 
dataset ``a''), $1.05\div1.30$ (low contamination - dataset ``b''), $>$1.30 
(significant contamination - dataset ``c'').
In each of the three cases, we applied the algorithm described in 
Sect.~\ref{recipe} (in particular, the steps corresponding to Sect. 
~\ref{merging}$\div$~\ref{analysis}) to extract and analyze the CXB 
spectrum, obtaining three independent determinations of the best fit CXB 
parameters.
The three total (CXB+quiescent NXB) spectra are 
plotted together in Fig.~\ref{spectra_differentnxb};
the corresponding best fitting normalizations and photon indexes  
are shown in Fig.~\ref{cont_spectra_differentnxb}.

It is evident that the parameters inferred from the most contaminated 
dataset (c) are 
totally incompatible with the results of dataset (a) and (b). The 
slope of the CXB spectrum is found to be flatter than expected, while 
the best fit normalization is higher by more than 100\% (more than 15$\sigma$ 
for a single interesting parameter) with respect to dataset (a). The results 
from datasets (a) 
and (b), on the contrary, are found to be in good agreement in both 
the slope and the normalization. Thus, the simple recipe described in 
Sect.~\ref{recipe} works well up to values of $\sim1.3$ for the ratio R, 
i.e. in presence of a low contaminating SP component.

As a further step we studied the spectral shape of the contaminating 
 background component.
The differences in count rate among the three datasets can be ascribed, in 
principle, to three causes: (i) presence of residual SP NXB, (ii) 
variations of the quiescent NXB, (iii) cosmic variance, due to the discrete 
nature of the CXB. 
In order to get the spectrum of the ``pure'' residual SP NXB, it is 
necessary to remove the variability due to the other causes.

As shown in Appendix ~\ref{qnxb}, the variation in the quiescent NXB level 
from observation to observation may be significant; however, merging several 
observations (at least six, as in dataset (c) for the MOS2), the variance 
should 
 be lower. We evaluated the actual level of the quiescent 
NXB in each of the three datasets using the corresponding {\em out FOV} 
spectrum. 
An 
{\em ad hoc} renormalized {\em closed} spectrum was then subtracted from each 
of 
the three spectra to remove the quiescent NXB component.

Cosmic variance should not play an important role in the 
comparison of the three spectra. As discussed by Barcons et al. (2000), 
the expected variance in the CXB intensity is of order 10\% on a solid 
angle $\Omega$ of 1 square degree, scaling roughly as 
$\Omega^{-{\frac{1}{2}}}$. The 
minimum solid angle is covered by dataset (c) ($\sim$1.5 square degree). 
Since
the CXB signal accounts only for $\sim10$\% of the 2-12 keV count rate, 
we see that only differences $<$1\% can be due to cosmic variance. 

Any difference among the three (quiescent NXB subtracted) spectra can 
therefore be ascribed to the residual SP NXB.
Assuming the level of such NXB component to be 0 in dataset (a), we extracted 
the pure residual SP NXB spectrum affecting datasets (b) and (c) by means of 
the simple subtraction of spectrum (a).


The NXB affecting dataset (c) is found to have a spectrum  well described 
by a power law having photon index of $\sim$-0.05 and an exponential cutoff at 
$\sim$5 keV, not convolved with the 
telescope effective areas and the CCD quantum efficiencies. This is the 
typical 
spectral shape of a particle-induced background component. Markevitch et 
al. (2002) reported a very similar result for the ``flare'' spectra affecting 
Chandra ACIS observations.

The good statistics allowed to repeat the analysis for each single 
observation belonging to dataset (c). The results clearly showed that the 
spectrum  of the  contaminating component is not stable. The significant 
differences in the spectral shape can be equally well reproduced, within the 
statistical uncertainty, with variations in the photon index or in the 
exponential cutoff position. We note that cosmic variance, even on a solid 
angle 
as small as the single observation field of view, could account only for a 
minor 
part of the observed variations.
It is very likely that the spectrum of the residual NXB component depend on 
the 
energy distribution of the impinging particles; in any case, more detailed 
analyses are beyond the scope of this work. 

Turning now to dataset (b), we find that the spectrum of the anomalous 
NXB, within the limits of a lower statistics, can be well described by the 
same 
model used for the case (c). 

\subsection{Conclusions}

The results on the study of the residual soft proton background can be 
summarized as follows:
\begin{itemize}
\item As a result of peculiar low-energy particle radiation conditions, a 
significant soft proton NXB component can survive after GTI screening. 
\item A study of the ratio of surface brightness $\Sigma_{IN}/\Sigma_{OUT}$ in 
the high energy range (8-12 keV) can easily reveal the presence of such 
residual 
NXB.
\item The spectral shape is a flat power law with an exponential cutoff, not 
convolved with the telescope 
effective areas and the CCD quantum efficiencies, typical of particle 
background.
\item The intensity can be highly variable. Increases in 
the NXB count rate up to 300\% can occur.
\item The spectral slope is highly variable and unpredictable, being possibly 
dependent on the energy distribution of the impinging particles.
\item Spectral study of extended sources may be heavily affected.
\item When the intensity of the residual soft proton NXB is low (up to 30\% 
higher than average) a 
simple renormalization of the 
quiescent background spectrum yields acceptable results owing to the very 
flat spectral shape of the low-level soft proton spectrum. The NXB level in 
the observation under exam must be evaluated within the FOV (better at 
large off-axis angles), using the high energy range (better excluding 
fluorescence lines, if the statistics is good).
\item There are hints for the presence of an irreducible flow of low-energy 
particles always reaching the detectors. Its impact on the analysis of the CXB 
is 
negligible.
\end{itemize}



\begin{acknowledgements}

We are grateful to all the members of the EPIC collaboration for their work and 
their support. We thank Alberto Moretti and Sergio Campana for providing their 
compilation of past CXB measurements.
We thank the referee, David Lumb, for his suggestions.
ADL acknowledges ASI for a fellowship.

\end{acknowledgements}

\end{document}